\documentclass[aps,prd,twocolumn,tightenlines,superscriptaddress,showpacs]{revtex4}
\newcommand{\taum} {\tau^{-}}
\newcommand{\epm} {e^{+}e^{-}}

\newcommand{\taupm} {\tau^{+}\tau^{-}}
\newcommand{\neutau} {\nu_{\tau}}
\newcommand{\piKs} { \pi^{-} K^0_S} 
 
\newcommand{\piKspizero} { \pi^{-} K^0_S \pi^{0} }
\newcommand{\piKsKs} { \pi^{-} K^0_S K^0_S }

\newcommand{\piKsKspizero} { \pi^{-} K^0_S K^0_S \pi^{0} }
\newcommand{\KKs} { K^{-} K^0_S }
\newcommand{\KKspizero} { K^{-} K^0_S \pi^{0} }

\newcommand{\KsX} { K^{0}_{S} ~ X^{-} } 
\newcommand{\Ks}  { K^{0}_{S} }

\newcommand{\Br}{\mathcal{B}} 
\newcommand{\tauTO} {\tau^{-} \rightarrow }

\newcommand{\Lum}{\mathcal{L}}

\newcommand{\piKsKl} { \pi^{-} K^0_S K^0_L }
\newcommand{\piKspizeropizero} { \pi^{-} K^0_{S} \pi^{0}\pi^{0} }
\newcommand{\piKseta} { \pi^{-} K^0_S \eta }
\newcommand{\piKsKlpizero} { \pi^{-} K^0_S K^0_L \pi^{0}}
\newcommand{\Kshhh} { K^0_S h^-h^+h^- }

\newcommand{\fone} { f_{1}} 
\newcommand{\foneprime}{f^{\prime}_{1}}
\newcommand{\Kstar}{K^{*-}}

\usepackage{graphicx} 
\usepackage{dcolumn}  
\usepackage{amsmath} 
\usepackage{amsfonts} 
\usepackage{amssymb}
\usepackage{hyperref} 
\usepackage{longtable}

\usepackage{graphicx,color} 
\definecolor{red}{rgb}{1,0,0}
\definecolor{green}{rgb}{0,1,0}

\begin{document}
\vspace*{-3\baselineskip}


\title{ \quad\\[1.0cm] Measurements of Branching Fractions of $\tau$ Lepton Decays with one or more $\Ks$}

\noaffiliation
\affiliation{University of the Basque Country UPV/EHU, 48080 Bilbao}
\affiliation{Beihang University, Beijing 100191}
\affiliation{Budker Institute of Nuclear Physics SB RAS and Novosibirsk State University, Novosibirsk 630090}
\affiliation{Faculty of Mathematics and Physics, Charles University, 121 16 Prague}
\affiliation{Deutsches Elektronen--Synchrotron, 22607 Hamburg}
\affiliation{Justus-Liebig-Universit\"at Gie\ss{}en, 35392 Gie\ss{}en}
\affiliation{Hanyang University, Seoul 133-791}
\affiliation{University of Hawaii, Honolulu, Hawaii 96822}
\affiliation{High Energy Accelerator Research Organization (KEK), Tsukuba 305-0801}
\affiliation{Indian Institute of Technology Guwahati, Assam 781039}
\affiliation{Indian Institute of Technology Madras, Chennai 600036}
\affiliation{Institute of High Energy Physics, Chinese Academy of Sciences, Beijing 100049}
\affiliation{Institute of High Energy Physics, Vienna 1050}
\affiliation{Institute for High Energy Physics, Protvino 142281}
\affiliation{INFN - Sezione di Torino, 10125 Torino}
\affiliation{Institute for Theoretical and Experimental Physics, Moscow 117218}
\affiliation{J. Stefan Institute, 1000 Ljubljana}
\affiliation{Kanagawa University, Yokohama 221-8686}
\affiliation{Institut f\"ur Experimentelle Kernphysik, Karlsruher Institut f\"ur Technologie, 76131 Karlsruhe}
\affiliation{Korea Institute of Science and Technology Information, Daejeon 305-806}
\affiliation{Korea University, Seoul 136-713}
\affiliation{Kyungpook National University, Daegu 702-701}
\affiliation{\'Ecole Polytechnique F\'ed\'erale de Lausanne (EPFL), Lausanne 1015}
\affiliation{Faculty of Mathematics and Physics, University of Ljubljana, 1000 Ljubljana}
\affiliation{Luther College, Decorah, Iowa 52101}
\affiliation{University of Maribor, 2000 Maribor}
\affiliation{Max-Planck-Institut f\"ur Physik, 80805 M\"unchen}
\affiliation{School of Physics, University of Melbourne, Victoria 3010}
\affiliation{Moscow Physical Engineering Institute, Moscow 115409}
\affiliation{Graduate School of Science, Nagoya University, Nagoya 464-8602}
\affiliation{Kobayashi-Maskawa Institute, Nagoya University, Nagoya 464-8602}
\affiliation{Nara Women's University, Nara 630-8506}
\affiliation{National United University, Miao Li 36003}
\affiliation{Department of Physics, National Taiwan University, Taipei 10617}
\affiliation{H. Niewodniczanski Institute of Nuclear Physics, Krakow 31-342}
\affiliation{Nippon Dental University, Niigata 951-8580}
\affiliation{Niigata University, Niigata 950-2181}
\affiliation{University of Nova Gorica, 5000 Nova Gorica}
\affiliation{Osaka City University, Osaka 558-8585}
\affiliation{Pacific Northwest National Laboratory, Richland, Washington 99352}
\affiliation{Panjab University, Chandigarh 160014}
\affiliation{University of Pittsburgh, Pittsburgh, Pennsylvania 15260}
\affiliation{University of Science and Technology of China, Hefei 230026}
\affiliation{Seoul National University, Seoul 151-742}
\affiliation{Soongsil University, Seoul 156-743}
\affiliation{Sungkyunkwan University, Suwon 440-746}
\affiliation{School of Physics, University of Sydney, NSW 2006}
\affiliation{Tata Institute of Fundamental Research, Mumbai 400005}
\affiliation{Excellence Cluster Universe, Technische Universit\"at M\"unchen, 85748 Garching}
\affiliation{Tohoku Gakuin University, Tagajo 985-8537}
\affiliation{Tohoku University, Sendai 980-8578}
\affiliation{Department of Physics, University of Tokyo, Tokyo 113-0033}
\affiliation{Tokyo Institute of Technology, Tokyo 152-8550}
\affiliation{Tokyo Metropolitan University, Tokyo 192-0397}
\affiliation{Tokyo University of Agriculture and Technology, Tokyo 184-8588}
\affiliation{University of Torino, 10124 Torino}
\affiliation{CNP, Virginia Polytechnic Institute and State University, Blacksburg, Virginia 24061}
\affiliation{Wayne State University, Detroit, Michigan 48202}
\affiliation{Yonsei University, Seoul 120-749}
  \author{S.~Ryu}\affiliation{Seoul National University, Seoul 151-742} 
  \author{I.~Adachi}\affiliation{High Energy Accelerator Research Organization (KEK), Tsukuba 305-0801} 
  \author{H.~Aihara}\affiliation{Department of Physics, University of Tokyo, Tokyo 113-0033} 
  \author{D.~M.~Asner}\affiliation{Pacific Northwest National Laboratory, Richland, Washington 99352} 
  \author{V.~Aulchenko}\affiliation{Budker Institute of Nuclear Physics SB RAS and Novosibirsk State University, Novosibirsk 630090} 
  \author{T.~Aushev}\affiliation{Institute for Theoretical and Experimental Physics, Moscow 117218} 
  \author{A.~M.~Bakich}\affiliation{School of Physics, University of Sydney, NSW 2006} 
  \author{A.~Bala}\affiliation{Panjab University, Chandigarh 160014} 
  \author{B.~Bhuyan}\affiliation{Indian Institute of Technology Guwahati, Assam 781039} 
  \author{A.~Bobrov}\affiliation{Budker Institute of Nuclear Physics SB RAS and Novosibirsk State University, Novosibirsk 630090} 
% ...
\collaboration{The Belle Collaboration}

\begin{abstract}
We report measurements of branching fractions of $\tau$ lepton decays to final states
with a $\Ks$ meson using a 669 fb$^{-1}$ 
data sample accumulated with the Belle detector at the 
KEKB asymmetric-energy $\epm$ collider.
The inclusive branching fraction is measured to be 
$\Br (\tau^{-} \to \Ks\ X^{-} \nu_{\tau})=(9.15 \pm 0.01 \pm 0.15) \times 10^{-3}$,
where $X^{-}$ can be anything; the exclusive branching fractions are
\begin{center}
$\Br(\taum \to \piKs \nu_{\tau}) = (4.16 \pm 0.01 \pm 0.08) \times 10^{-3},$ \\
$\Br(\taum \to \KKs \nu_{\tau}) = (7.40 \pm 0.07 \pm 0.27) \times 10^{-4},$  \\
$\Br(\taum \to \piKspizero \nu_{\tau}) = (1.93 \pm 0.02 \pm 0.07) \times 10^{-3},$\\  
$\Br( \taum \to \KKspizero \nu_{\tau}) = (7.48 \pm 0.10 \pm 0.37)\times 10^{-4},$  \\ 
$\Br( \taum \to \pi^{-} \Ks \Ks \nu_{\tau}) = (2.33 \pm 0.03 \pm 0.09) \times 10^{-4},$ \\  
$\Br( \taum \to \piKsKspizero \nu_{\tau}) = (2.00 \pm 0.22 \pm 0.20) \times 10^{-5},$
\end{center}
where the first uncertainty is statistical and the second is systematic.
For each mode, the accuracy is improved over that of pre-$B$-factory measurements 
by a factor ranging from five to ten.
In $\tauTO \piKsKspizero \nu_{\tau}$ decays, 
clear signals for the intermediate states  
$\tauTO \pi^-\fone(1285)\nu_{\tau}$ and  
$\tauTO K^{*-}\Ks \pi^{0} \nu_{\tau}$
are observed.

\end{abstract}

\pacs{13.25.-k, 14.60.Fg, 13.35.Dx}

\maketitle

\tighten

{\renewcommand{\thefootnote}{\fnsymbol{footnote}}}
\setcounter{footnote}{0}

\section{Introduction}
\label{sec:introduction}

Hadronic $\tau$ decays provide a clean environment for the study of low-energy 
hadronic currents. In these decays, the hadronic system is produced 
from the QCD vacuum via the
charged weak current mediated by a $W$ boson. The $\tau$ decay amplitude 
can thus be factorized into a purely leptonic part including $\tau$ and 
$\nu_{\tau}$ and a
hadronic spectral function that measures the transition probability 
to create hadrons out of the vacuum.
The Cabibbo-favored (non-strange) spectral function measured in the ALEPH 
and OPAL experiments has been used for 
detailed QCD studies and resulted in a precise determination of the 
strong coupling constant 
$\alpha_{s}(M_{z}^{2})$~\cite{Narison:1988ni, Schael:2005am, Ackerstaff:1998yj}.

Decays of $\tau$ leptons to final states containing one or more $\Ks$ mesons are 
of importance in order to address issues in both Cabibbo-favored (non-strange) 
and Cabibbo-suppressed (strange) spectral functions. 
In particular, by studying decays into final states 
that contain an odd number of kaons, 
one can extract the strange spectral functions and determine the 
Cabibbo-Kobayashi-Maskawa (CKM) matrix element 
$|V_{us}|$~\cite{Gamiz:2004ar, Baikov:2004tk, Kambor:2000dj}. 
On the other hand, modes with an even number of kaons play an important role 
in understanding the non-strange vector and axial-vector components. 
Precision measurements of the branching fractions for various processes are 
essential for these studies.

Despite extensive studies of $\tau$ hadronic decays performed at LEP 
and CLEO, prior to the $B$ factory era, Cabibbo 
and phase-space suppression have resulted in limited statistics for the studies of kaon production in hadronic $\tau$ decays~\cite{Barate:1999hi, Barate:1999hj, Abbiendi:1999pm, PhysRevD.53.6037}.

Experiments at the $B$ factories have provided improved measurements of the 
branching fractions and spectral functions for modes with kaons:
$\tau^-\to \piKs \nu_{\tau}$~\cite{Epifanov:2007rf},  
 $\tau^-\to K^-\pi^{0} \nu_{\tau}$~\cite{Aubert:2007jh},  
three charged hadrons~\cite{Lee:2010tc, Aubert:2009qj, Aubert:2007mh} and
modes that include an $\eta$ meson~\cite{Inami:2008ar, delAmoSanchez:2010pc}.
(Unless otherwise specified, charged-conjugate decay modes are implied throughout this paper.)
Recently, the BaBar collaboration reported an improved branching fraction 
and a first measurement for the rare decay processes 
$\taum \to \piKsKs \nu_{\tau}$ and 
$\taum \to \piKsKspizero \nu_{\tau}$, respectively~\cite{Lees:2012de}.

In this article, we report precision measurements of the branching fractions of $\tau$ lepton decays 
for the inclusive and various exclusive modes with $\Ks$ mesons in the final state. 
The $\Ks\to \pi^+\pi^-$ decay is used for the $\Ks$ meson reconstruction.
We measure the inclusive branching fraction for $\tau^{-} \to \Ks\ X^{-} \nu_{\tau}$, 
where $X^{-}$ stands for anything, from the final state that containing $\Ks$ mesons 
in the sample.
The candidates are then classified according to the number of $\Ks$ mesons,
as well as the numbers of $\pi^{0}$, $\pi^{-}$ and $K^{-}$ mesons. 
We use these sorted events to measure the exclusive branching fractions 
for the following six modes:
\begin{eqnarray}
\taum \to \piKs \nu_{\tau}, \nonumber \\
\taum \to \KKs \nu_{\tau}, \nonumber \\
\taum \to \piKspizero \nu_{\tau}, \nonumber\\
\taum \to \KKspizero \nu_{\tau}, \nonumber\\
\taum \to \piKsKs \nu_{\tau},  \nonumber\\
\taum \to \piKsKspizero \nu_{\tau}.  \nonumber
\end{eqnarray}
Since some modes are the main source of the backgrounds for other modes,
we measure the branching fraction of these six modes simultaneously
by means of an efficiency matrix.

\section{Data set, detector and data modeling} 
\label{sec:detector}

The present analysis uses a data sample of 669 fb$^{-1}$ collected 
with the Belle detector at the KEKB asymmetric-energy $e^+e^-$ 
collider~\cite{Kurokawa:2003io,Abe2013aa} running on the 
$\Upsilon(4S)$ resonance, 10.58 GeV, and 60 MeV below it (off-resonance).
This sample contains 616$\times 10^6$ $\taupm$ pairs, which is two 
orders of magnitude larger than those that were available prior 
to the $B$-factory experiments.
The Belle detector is a large-solid-angle magnetic spectrometer 
that consists of a silicon vertex detector (SVD), a 50-layer central 
drift chamber (CDC), an array of 1188 aerogel threshold Cherenkov 
counters (ACC), 
a barrel-like arrangement of time-of-flight scintillation counters (TOF), 
and an electromagnetic calorimeter (ECL) comprised 
of 8736 CsI(Tl) crystals located inside a superconducting solenoid 
coil that provides a 1.5~T magnetic field. 
An iron flux return located outside of the coil is instrumented to 
detect $K_L^0$ mesons and to identify muons (KLM).
The detector solenoid is oriented along the $z$ axis, 
pointing in the direction opposite that of the positron beam.
The $r-\phi$ plane is transverse to this axis.

Two inner detector configurations are used in this analysis.
A beam pipe with a radius of 2.0 cm and a 3-layer silicon vertex detector 
are used for the first sample of $142 \times 10^{6}$ $\taupm$ pairs, 
while a 1.5 cm beampipe, a 4-layer silicon detector and a small-cell inner 
drift chamber are used to record 
the remaining $474 \times 10^{6}$ $\taupm$ pairs~\cite{Natkaniec:2006rv}.
The detector is described in detail 
elsewhere~\cite{Abashian2002117,Brodzicka2012aa}.

The {\tt KKMC}~\cite{Jadach:1999vf} code is used to generate
the $\tau$-pair production $e^+e^-\to \tau^+\tau^- (\gamma)$, 
and the {\tt TAUOLA}/{\tt PHOTOS}~\cite{Jadach:1993hs,Golonka:2003xt}
codes to describe the $\tau$ lepton decays.
The values of the branching fractions in these codes are updated 
to the recent measurements reported in Ref.~\cite{Beringer:1900zz}. 

The generated events are then passed through a full detector
simulation based on {\tt GEANT}~\cite{Brun:1987ma} and the same analysis 
program as used for the data.
The efficiencies of the reconstruction of charged tracks and $\pi^{0}$ and 
of particle identification (PID)
are calibrated with data and corrections are applied to the Monte Carlo (MC) results as 
discussed in Section \ref{sec:backgroundandefficiency}.

The background from non-$\tau$ events from continuum $e^+e^-\to q \bar{q}$ 
(where $q=u,d,s,c$), $B\bar{B}$ and two-photon events is modeled with the 
{\tt JETSET}~\cite{Sjostrand:1993yb}, 
{\tt EVTGEN}~\cite{Lange:2001uf} and {\tt AAFH}~\cite{Berends:1986ig} codes, respectively.
 

\section{Event Selection and reconstruction} 
\label{sec:eventselection}
The selection process, which is optimized to suppress background 
while retaining high efficiency for the decays under study, 
proceeds in two stages: the selection of $\epm \to \taupm$ events
and the extraction of events that contain one or more $\Ks$ mesons.

\subsection{Selection of $\taupm$ pair events}
The $\tau$-pair selection is focused on suppressing other 
physical processes as well as keeping single-beam 
induced background at a negligible level.
Loose conditions are applied for $\tau$-pair selection 
in terms of the number of charged tracks.
We select events having at least two and as many as six tracks 
with a net charge equal to zero or $\pm1$.
Each track is required to have a momentum transverse to the beam axis ($p_{\rm T}$) 
greater than 0.1 GeV$/c$.
It must have a distance of closest approach to the interaction point (IP) within $\pm$ 3.0 cm 
along the beam direction (the $z$ axis) and 1 cm in the transverse ($r$--$\phi$) plane.
We include tracks that fail the IP condition if they are daughters of $\Ks$ candidates.
(Most $\Ks$ daughters satisfy the IP requirement.)
We also perform a vertex fit of the tracks satisfying the IP requirement
and require the primary vertex position 
to be within $\pm$ 3.0 cm along the $z$ axis and 0.5 cm in the $r$--$\phi$ plane.

Each photon (reconstructed from a cluster in the calorimeter) must be
separated from the nearest track projection by at least 20 cm.
The energy of each photon must be greater than 80~MeV
in the barrel region ($31^{\circ}< \theta < 128^{\circ}$) and
greater than 100~MeV in the endcap regions ($17^{\circ}< \theta < 30^{\circ}$
and $130^{\circ}< \theta < 150^{\circ}$), where $\theta$ is the polar angle 
with respect to the $z$ axis in the laboratory frame.
The sum in the center-of-mass (CM) frame of the 
magnitudes of the track momenta and 
the energies of all photon candidates must be less than 9 GeV.
Backgrounds from two-photon and QED $\epm \to \ell^{+}\ell^{-}$ processes, 
where $\ell$ is an electron or muon, are reduced by requiring the missing mass, 
$M_{\rm miss} = \sqrt{p^{2}_{\rm miss}}$ 
(where $p_{\rm miss} = p_{\rm init}-\sum_{i} p_{{\rm track},i}-\sum_{i} p_{\gamma,i}$),
and the polar angle of the missing momentum in the CM frame  
to satisfy 1~GeV$/c^{2}$ $< M_{\rm miss}<$ 7~GeV$/c^{2}$ and
$30^{\circ} < \theta_{\rm miss} <150^{\circ}$.
In the definition of the missing mass, $p_{ {\rm track},i}$ and $p_{\gamma,i}$ are
the four-momentum of the $i$-th track and photon, respectively,
and $p_{\rm init}$ is the initial four-momentum of the colliding $e^+e^-$ system.
The pion mass is assigned to all of the measured tracks 
that are not identified as electrons or muons.

The $\tau$ pairs are produced back-to-back in the $e^+e^-$ CM frame. 
As a result, the decay products of the two $\tau$ leptons can be separated 
from each other by dividing the event into two hemispheres.
The hemispheres are defined in the CM by the plane perpendicular to the thrust axis $\hat{n}$, 
defined as the unit vector in the direction of the thrust 
$T = {\rm max} \left[ \Sigma_{i}| \hat{n} \cdot \vec{p_{i}}| / \Sigma_{i}|\vec{p}_{i} | \right]$,
where $\vec{p}_{i}$ is the momentum of the $i$-th particle, either a track or a photon.
Each event is required to have exactly one track in one of the hemispheres (tag side) 
and one or more tracks in the other hemisphere (signal side).
The continuum background ($e^{+}e^{-} \to q\bar{q}$) is suppressed by requiring $T > 0.9$.
In addition, the sum of the charges of all tracks should 
vanish, where we include here the daughters of all $\Ks$ candidates.
This requirement reduces background further while sustaining the efficiency of
$\Ks$ candidates with a relatively long flight length.

Particle identification for charged tracks is crucial in this analysis.
Information from several subsystems is used to identify the type of 
charged particle: electron, muon, pion and kaon.
For lepton identification,  we form likelihoods 
$\mathcal{L}(e)$ for electron \cite{Hanagaki:2001fz}
and $\mathcal{L}(\mu)$ for muon \cite{Abashian:2002bd} 
using the response of the appropriate sub-detectors.
An electron track is clearly identified from  the ratio of the energy deposited 
in the electromagnetic calorimeter to the momentum measured in the tracking 
subsystems (the $E/p$ ratio) and the shower shape in the ECL at high momentum
and from the $dE/dx$ information measured in the CDC at low momentum. 
We require $\mathcal{L}(e) > 0.9$ to identify electrons.
Under these conditions, the efficiency is greater than 95\% and the fake rate is less than 1\%.
A muon track is identified mainly from the range and transverse scattering in the  KLM detector.  
We require $\mathcal{L}(\mu) > 0.9$ and a momentum greater than 0.7 GeV$/c$.
The efficiency is greater than 95\% and the fake rate is less than 3\% for particles with momenta 
above 1.0 GeV$/c$. 

To distinguish hadron species, we use a likelihood ratio 
$\mathcal{L}(i/j) = \mathcal{L}_{i}/( \mathcal{L}_{i} + \mathcal{L}_{j})$, 
where $\mathcal{L}_{i}\ (\mathcal{L}_{j})$ is 
the likelihood of the detector response for a particle of type $i$ ($j$).
For separation of charged pions and kaons, the hit information from 
the ACC, the $dE/dx$ information in the CDC, 
and the time-of-flight are used.
On the signal side, a track not identified as either an electron or a muon  
is identified as a kaon (pion) when $\mathcal{L}(K|\pi) > 0.7 (<0.7)$.
The kaon and pion identification efficiencies are typically
$83-85\%$ and $93-95\%$, respectively. 
The probabilities to misidentify a pion as a kaon and a kaon as a pion 
are in the range $5-7$\% and $15-17$\%, respectively.

Events useful for this analysis are classified in the following three categories
according to the contents of the signal side:
1) one $\Ks$, 2) two $\Ks$ and 3) one lepton. 
For categories 1 and 3, the tag side contains one lepton, while 
category 2 requires one charged track in the tag side.
The third category, with two leptons, is used for the normalization of the 
branching fraction measurements.

\begin{figure}
\begin{center}
\includegraphics[width=0.95\columnwidth]{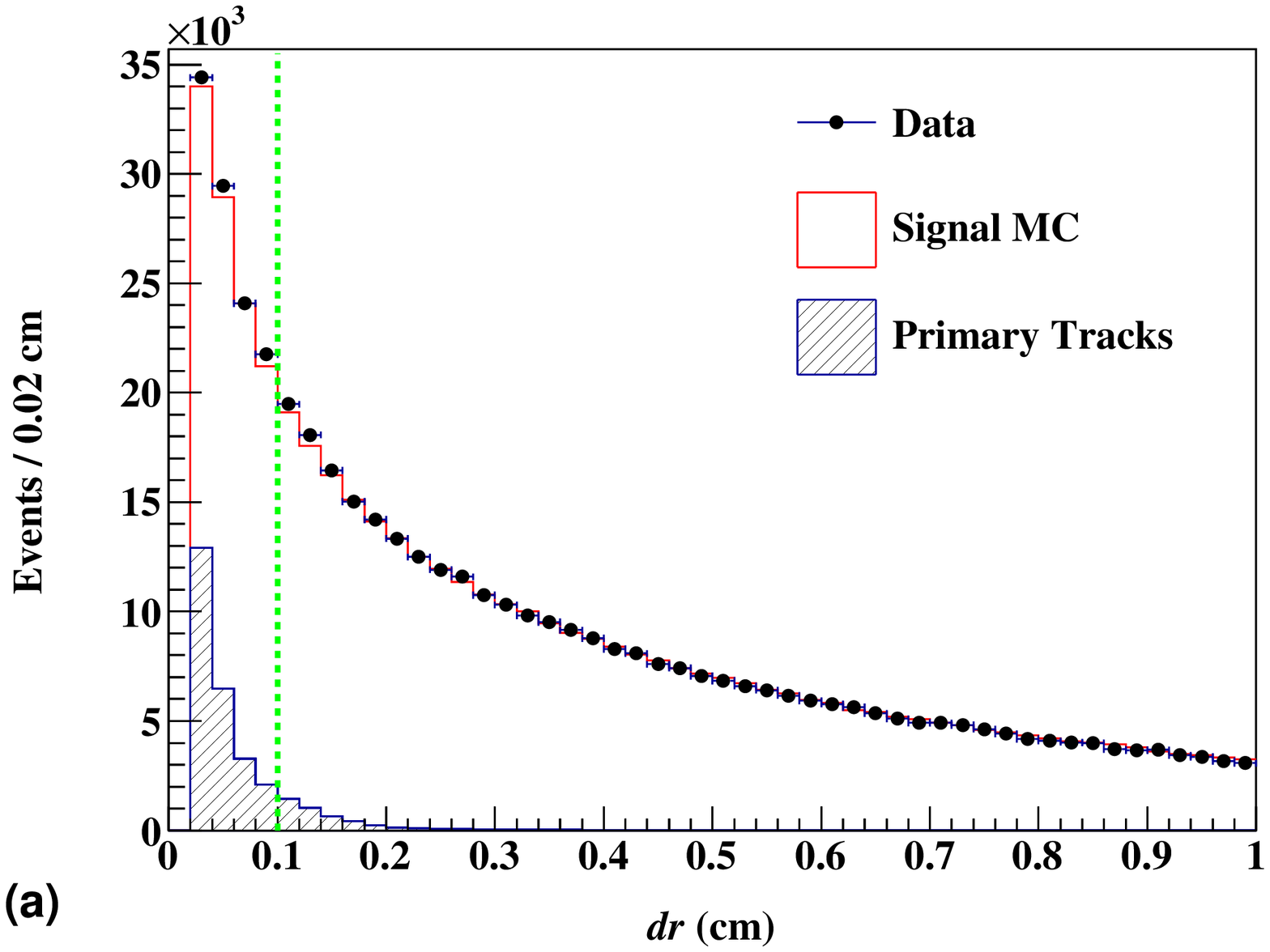}
\includegraphics[width=0.95\columnwidth]{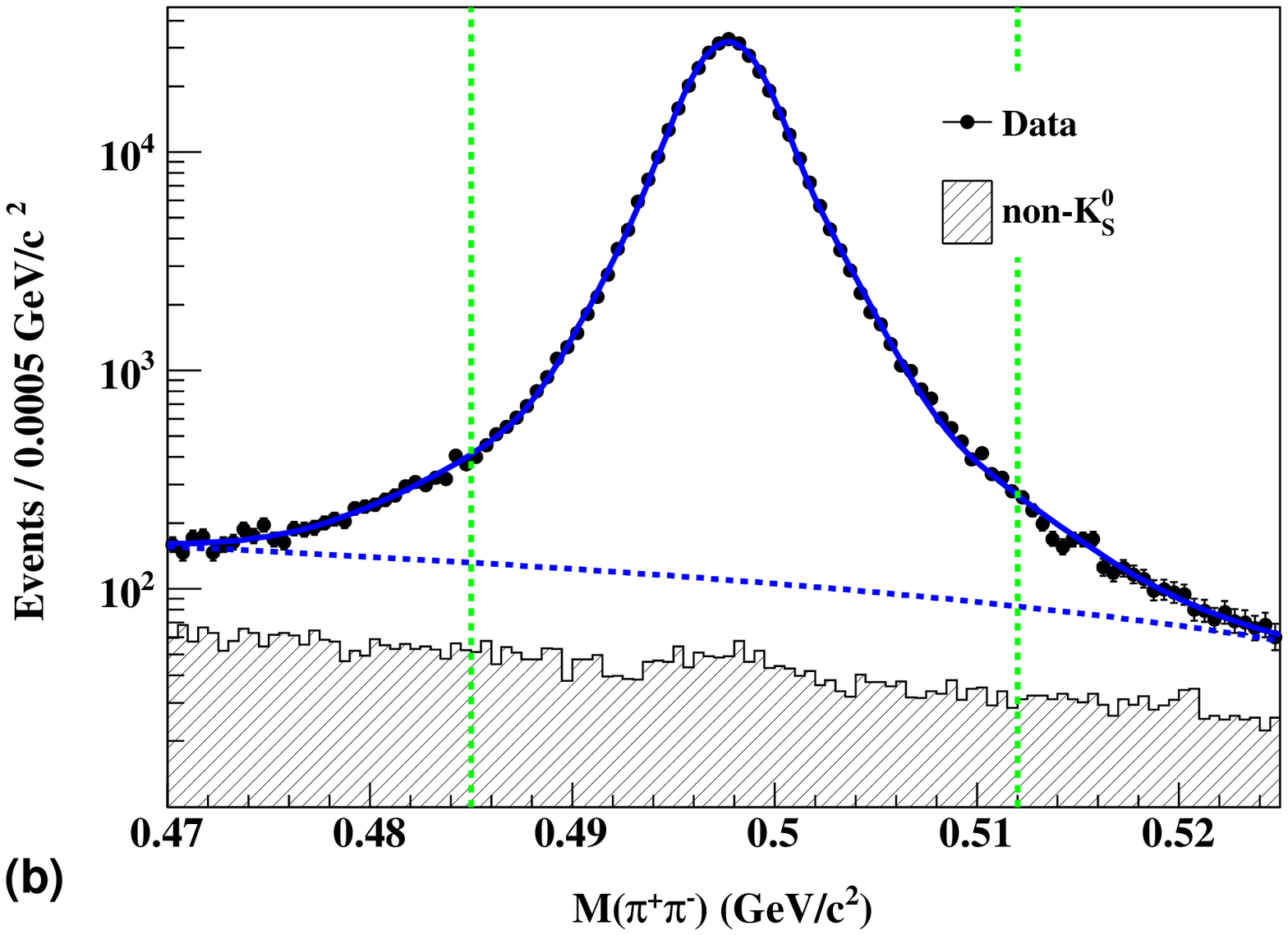}
\end{center}
\caption{(color online) Selection of $\Ks \to \pi^{+} \pi^{-}$ 
candidates in $\tau$ decays: 
(a) Distribution of the closest distance of approach 
to the IP in the $r-\phi$ plane for two $\Ks$ daughter tracks of the $\Ks$ candidates.
The background represented by the shaded histogram, obtained by MC, consists  
of the tracks from primary vertex.
(b) Distribution of the invariant mass $M(\pi^{+}\pi^{-})$ for the $\Ks$ candidates 
after applying all selection requirements except the mass.
The solid line is a fit with three Gaussians for the signal and a linear 
background. 
The shaded histogram stands for the background from 
$\tau^{-} \to \pi^{-}\pi^{+}\pi^{-} \nu_{\tau}$ obtained by MC.
In both plots, the vertical lines represent the $\Ks$ selection criteria.
}
\label{fig:ksreconst}
\end{figure}

\subsection{Selection of events containing one $\Ks$}
\label{subsec:selectionofeventscontainingoneks}

For the modes with one $\Ks$, the $\Ks \to \pi^+\pi^-$ candidate 
is reconstructed from a pair of oppositely charged tracks.
The $z$-distance between the two helices at the $\pi^{+}\pi^{-}$ vertex 
position ($z_{\rm dist}$) must be less than 2.5 cm.
The pion momenta are then refitted with a common vertex constraint.  
The flight length ($\ell_{f}$) of the $\Ks$ candidates must be between 2 cm and 20 cm.  
The distance of closest approach to the IP in the 
$r$--$\phi$ plane ($dr$) is required to be larger than 0.1 cm 
for each daughter in order to suppress the 
background from the tracks from the primary vertex.
The $dr$ distribution is well reproduced by MC as shown 
in Fig.~\ref{fig:ksreconst} (a). 
Figure~\ref{fig:ksreconst} (b) shows the  distribution of the 
$\pi^+\pi^-$ invariant mass of the $\Ks$ candidates.
A clear $\Ks$ signal is seen with a small background that is less than 1\%.
The signal window is defined as the mass range 
0.485 GeV/$c^{2} < M_{\pi\pi} < 0.512$ GeV/$c^{2}$,
which corresponds to a $\pm 5\sigma$ window.

Events containing at least one so-defined $\Ks$ are assigned to the inclusive $\Ks$ 
sample irrespective of the accompanying particles on the signal side.
The number of inclusive $\tau^-\to \KsX \nu_{\tau}$ events in this sample 
is obtained from a fit to the $\pi^+\pi^-$ invariant mass distribution that uses 
the sum of three Gaussians for the signal and a linear function for background. 
In the case where an event contains two or more $K^0_S$ candidates,
one is chosen arbitrarily for the fit.
The fit, shown as the solid curve in Fig.~\ref{fig:ksreconst} (b), 
yields $397806\pm 631$ inclusive $\Ks$ signal events.

\begin{figure}[t]
\begin{center}
\includegraphics[width=0.95\columnwidth]{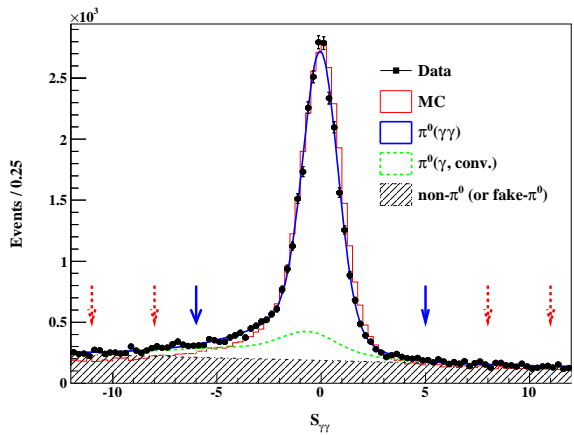}
\end{center}
\caption{ (color online)
Distribution of the normalized two-photon invariant mass
$S_{\gamma\gamma}=(M_{\gamma\gamma} - M_{\pi^{0}})/\sigma_{\gamma\gamma}$ 
for $\tau^{-} \to \piKspizero \nu_{\tau}$ candidates.
The arrows indicate the signal and sideband region.
The area enclosed by the solid and dotted curves represents the true $\pi^{0}$ events 
reconstructed with two unconverted photons, the area enclosed by the dotted and the 
dot-dashed curves represents the true $\pi^{0}$ events reconstructed with
at least one converted photon, while the hatched area indicates the fake $\pi^{0}$ background events.
}
\label{fig:pi0reconst}
\end{figure}

For the modes with one $\pi^{0}$, $\piKspizero \nu_{\tau}$ and
$\KKspizero \nu_{\tau}$, the $\pi^{0}$ candidate is reconstructed from the
invariant mass of two photons detected on the signal side.
The normalized mass difference between the invariant mass of the two
photons and the nominal $\pi^{0}$ mass ($M_{\pi^{0}}$),
\begin{align}
S_{\gamma\gamma} = (M_{\gamma\gamma}-M_{\pi^{0}})/\sigma_{\gamma\gamma},
\end{align}
(where $\sigma_{\gamma\gamma}$ is the resolution of the invariant mass of the 
two photons) is used to determine the number of genuine $\pi^{0}$'s and to 
estimate the level of background from sidebands.
The value of $\sigma_{\gamma\gamma}$ ranges from 0.004 to 0.009~GeV/$c^{2}$, 
depending on the momentum and polar angle of the $\pi^{0}$ candidate.
The $S_{\gamma\gamma}$ distribution for events with one charged track and one 
$\Ks$ is shown in Fig.~\ref{fig:pi0reconst}.
The lower-side tail of the $S_{\gamma\gamma}$ distribution is primarily due 
to leakage of electromagnetic showers out of the CsI(Tl) 
crystals and the conversion of photons in the material located in front of 
the crystals.
Good agreement between data and MC indicates that
these effects are properly modeled by the MC simulation.
We define the interval $-6< S_{\gamma\gamma}< 5$ as the $\pi^0$ signal region.
We also use both  sideband regions, $8<|S_{\gamma\gamma}|<11$, 
for the estimation of the spurious $\pi^{0}$ background.
The sideband subtraction effectively removes the contamination of the 
spurious $\pi^{0}$ background in the selected samples.

As an alternative method, we count the number of the $\pi^{0}$ signal events
by fitting the $S_{\gamma\gamma}$ distribution with the following formula:
\begin{align}
f(S_{\gamma\gamma}) = N_{\pi^{0}} ((1- \alpha) S_{1}(S_{\gamma\gamma}) + \alpha S_{2}(S_{\gamma\gamma}))  \notag \\
                    + N_{\textrm{bg}} B(S_{\gamma\gamma}),
\end{align}
where $N_\pi^{0}$ and $N_{\textrm{bg}}$ are the yields of $\pi^{0}$ signal 
and non-$\pi^{0}$ background, respectively. 
$S_{1}(S_{\gamma\gamma})$ is the $\pi^{0}$ signal probability density function (PDF) 
where both photons from $\pi^{0}$ are detected by the ECL directly, 
while $S_{2}(S_{\gamma\gamma})$ is the $\pi^{0}$ signal PDF where at least 
one photon is converted by the material in front of the ECL. 
$B(S_{\gamma\gamma})$ is the PDF for non-$\pi^{0}$ background.
The shapes of $S_{1}(S_{\gamma\gamma})$, $S_{2}(S_{\gamma\gamma})$ and $B(S_{\gamma\gamma})$ 
are obtained from the MC simulation and are parametrized with a logarithmic Gaussian for
 $S_{1}(S_{\gamma\gamma})$ and $S_{2}(S_{\gamma\gamma})$ and a linear
function for $B(S_{\gamma\gamma})$.
The functional form of the logarithmic Gaussian is given in Appendix.
The parameter $\alpha$ is the probability that at least one $\gamma$ is converted.
In the fit to the data, the value of $\alpha$ is fixed to the MC value.
The fit results for the $S_{1}(S_{\gamma\gamma})$, 
$S_{2}(S_{\gamma\gamma})$, and $B(S_{\gamma\gamma})$ components 
are shown in Fig.~\ref{fig:pi0reconst}. 
The area enclosed by the solid  and dotted curves
represents the signal $S_{1}(S_{\gamma\gamma})$ component,
the area enclosed by the dotted and dot-dashed curves represents
the $S_{2}(S_{\gamma\gamma})$ component, and the hatched area indicates
the fake $\pi^{0}$ background.
The $S_{2}(S_{\gamma\gamma})$ component has a tail in the lower 
$S_{\gamma\gamma}$ region, 
since part of the $\gamma$ energy is lost by the conversion.
We obtain consistent results for the branching fraction for both methods and 
assign the difference, if any, as a systematic error.

The inclusive $\Ks$ sample is further subdivided into exclusive modes 
according to the number of $\Ks$ mesons, 
the number of charged hadrons and the number of $\pi^0$'s as: 
$\piKs \nu_{\tau}$, $\KKs \nu_{\tau}$, $\piKspizero \nu_{\tau}$ and 
$\KKspizero \nu_{\tau}$.
In order to determine the exclusive decay mode and reduce the contribution 
from decay modes with multiple $\pi^{0}$'s, the sum of the energies of any photons 
that are located on the signal side and not used for 
the $\pi^{0}$ reconstruction is required to be smaller than 0.2 GeV for all modes.
Finally, 157836 $\piKs\nu_{\tau}$, 32701 $\KKs \nu_{\tau}$, 
26605 $\piKspizero\nu_{\tau}$ and 8267 $\KKspizero\nu_{\tau}$ candidates 
are selected.

\begin{figure}
\includegraphics[width=0.95\columnwidth]{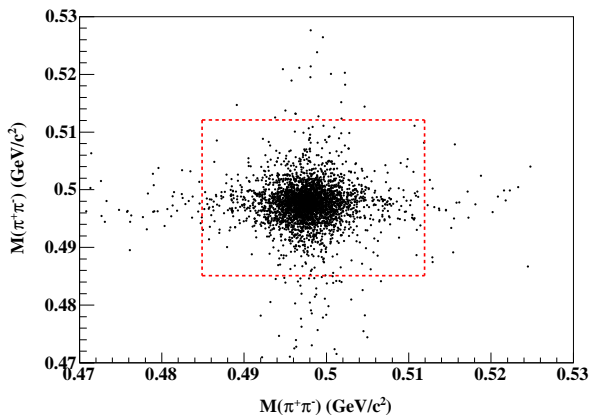}
\caption{
Two dimensional distribution of the $M(\pi^+\pi^-)$ invariant masses
for $\Ks\Ks$ candidate events in the $\tau$ decays. 
The dotted box is the $\Ks\Ks$ signal region.}
\label{fig:doubleksreconst}
\end{figure}

The selected number of events, as well as the background and selection 
efficiency discussed below, are summarized in Table~\ref{tab:selection}.

\subsection{Selection of events containing two $\Ks$ mesons}

Since low branching fractions ($\mathcal{O}(10^{-4})$ -- $\mathcal{O}(10^{-5})$) 
are expected for the modes with two $\Ks$ mesons,
several selection criteria are somewhat loosened for both $\Ks$ candidates
compared to those used to select single $\Ks$ events in order to increase the signal efficiency.
For the selection of $\Ks$, the criteria for $dr$, $\ell_{f}$ and $z_{dist}$ are 
$dr>0.01\ \rm{cm}$, $\ell_{f} < 50\ \rm{cm}$ and $z_{\rm dist} < 3.5\ \rm{cm}$. 
In addition, the requirements for the tag side are loosened so that there
is one charged track and any number of photons. No particle identification is
required for the charged track. 

Figure~\ref{fig:doubleksreconst} shows the two-dimensional invariant mass 
of the $\Ks\Ks$ candidates; a clear $\Ks\Ks$ signal is seen with negligible background. 
The signal is selected within the signal box 
$0.485~\textrm{GeV}/c^{2} < M (\pi^{+}\pi^{-}) < 0.512~\textrm{GeV}/c^{2}$, 
corresponding to a $\pm 5 \sigma$ window. 
The $\tauTO \piKsKs(\pi^{0}) \nu_{\tau}$ signal candidates are then selected 
with the condition of one $\pi^-$ and two $\Ks$ (plus one $\pi^{0}$).
Moreover, events where the energy sum of extra photons exceeds 0.3 GeV 
on the signal side are rejected.
Finally, 6684 $\piKsKs\nu_{\tau}$ and 303 $\piKsKspizero$ candidates 
are selected (summarized in Table~\ref{tab:selection}).

\subsection{Selection of two-lepton events}

The two-lepton events where both $\tau$ leptons decay leptonically 
are used for the normalization of the branching fraction measurements.
Only events with two leptons of different flavors (one electron and one muon)
are used, since di-electron and di-muon events are contaminated by the radiative 
Bhabha and $e^{+}e^{-} \to \mu^{+}\mu^{-}(\gamma)$ processes, respectively. 
We require the opening angle between the two leptons to exceed $90^{\circ}$ in the CM. 
This procedure selects 7.66 $\times 10^{6}$ $e-\mu$ events. 
 
\begin{figure}
\includegraphics[width=0.95\columnwidth]{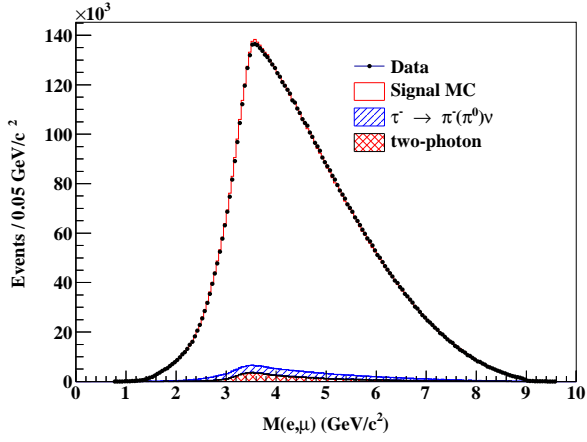}
\caption{(color online) Distribution of $e-\mu$ invariant mass.
The closed circles are data and the histogram is the sum of the signal
and background in the MC.
The hatched region and cross-hatched regions are the 
contributions from $\tau^{-} \to \pi^{-}\pi^{0}\nu_{\tau}$ and two-photon
processes, respectively.
} 
\label{fig:emu_mass}
\end{figure}

A detailed study using simulated data indicates that the background comes 
from the two-photon process 
$e^{+}e^{-} \to e^{+}e^{-}\mu^{+}\mu^{-}(\gamma)$ (1.6\%) and one-prong 
$\tau$ decay with leptonic $\tau$ decay on the other side, 
$\tau^{-} \to h^{-}(\pi^{0})\nu_{\tau}$, 
where $h^{-}=(\pi^-$, $K^-)$ is misidentified as a lepton (2.6\%).
The total background fraction in selected events is found to be 4.2\%. 
The detection efficiency is (19.31 $\pm$ 0.03)\%.
A comparison of the $e-\mu$ invariant mass distribution for 
data and MC, shown in Fig.~\ref{fig:emu_mass}, indicates good agreement
and that the performance of the detector is well understood.
In addition, the total number of $e-\mu$ events agrees within 0.38\%
with the expected number of events obtained from the integrated luminosity, 
the $\tau$-pair cross section, and the leptonic $\tau$ branching fractions.
This result is consistent with the uncertainty estimated in the 
luminosity measurement.  

\begin{table*}
\caption{ Results of the event selection.
Total efficiency ($\epsilon$),  number of selected events ($N^{\rm Data}$), background fraction ($R^{\rm Bg} = N^{\rm Bg}/N^{\rm Data}$), 
and number of signal events after 
background subtraction and efficiency correction ($N^{\rm Sig}$). 
The $\epsilon$ includes the $\Ks \to \pi^{+}\pi^{-}$ branching fraction.
}
\label{tab:selection}
\begin{tabular}{c c c c c  } \toprule
Decay mode & $\epsilon$ (\%) & $N^{\rm Data}$ &  $R^{\rm Bg} (\%) $ & $N^{\rm Sig}$   \\ \colrule
$\KsX$       & 9.66 & $397806 \pm 631$ & $4.20 \pm 0.46$  & $(3.947 \pm 0.007)\times 10^{6}$ \\  
$\piKs$      & 7.09 & $157836 \pm 541$ & $8.86 \pm 0.05$  & $(1.793 \pm 0.005)\times 10^{6}$ \\      
$\KKs$       & 6.69 & $32701 \pm 295$  & $3.55 \pm 0.07$  & $(3.193 \pm 0.018)\times 10^{5}$ \\     
$\piKspizero$& 2.65 &$26605\pm 208$    & $5.60 \pm 0.10$  & $(8.336 \pm 0.070)\times 10^{5}$ \\     
$\KKspizero$ & 2.19 & $8267\pm 109$    & $2.43 \pm 0.10$  & $(3.226 \pm 0.045)\times 10^{5}$  \\         
$\piKsKs$    & 2.47 &  $6684\pm 96$    & $7.89 \pm 0.24$  & $(2.447 \pm 0.033)\times 10^{5}$  \\          
$\piKsKspizero$ & 0.82 & $303 \pm 33$  & $11.60 \pm 1.60$  & $(2.105\pm 0.140)\times 10^{4}$  \\ \botrule   
\end{tabular}
\end{table*}

\section{Determination of the branching fractions}

\subsection{Formula for branching fraction measurements}
\label{sec:exclusivebr}
We use two different normalization methods for the determination of the 
branching fractions: one uses the number of $e-\mu$ events while the other 
uses the integrated luminosity.
As the number of $e-\mu$ events obtained from $\tau$-pair selection is 
consistent with the one deduced from the integrated luminosity measurement 
within 0.38\%, the normalization using either of them will lead 
to consistent results.
However, the resulting systematic uncertainties for the branching fraction 
measurements differ.

In the first method, the branching fraction is given by the formula
\begin{align}
\mathcal{B}_{i} = \frac{ N^{\rm Sig}_{i }} {N^{\rm Sig}_{e \textrm{-}\mu}} \frac{\mathcal{B}_{e} \mathcal{B}_{\mu}} {\mathcal{B}_{e}+\mathcal{B}_{\mu}},
\label{eq:bfltag}
\end{align}
where $i$ represents the decay mode under study and $N^{\rm Sig}_{i}$ is 
the number of signal events after efficiency and background corrections,
where one $\tau$ lepton decays into a signal mode and the other $\tau$ decays
leptonically. 
$N^{\rm Sig}_{e \textrm{-} \mu}$ is the number of $e-\mu$ events 
after efficiency and background corrections.
$\Br_{e}$ and $\Br_{\mu}$ are the branching fractions for 
$\tau^{-} \to e^{-} \bar{\nu}_{e} \nu_{\tau}$ and $\tau^{-} \to \mu^{-} \bar{\nu}_{\mu} \nu_{\tau}$, respectively.
The world-average values, 
$\Br_{e} = (17.83\pm0.04)\%$ and 
$\Br_{\mu} = (17.41\pm0.04)\%$~\cite{Beringer:1900zz},
are used. 
In this formula, the systematics coming from the luminosity measurement, 
tracking efficiency and the particle identification 
efficiency cancel (completely or partially) in the ratio. 
The branching fractions for the inclusive $\tau^{-} \to \Ks X^{-} \nu_{\tau}$
and four exclusive decay modes, 
$\tau^-\to \piKs \nu_{\tau}$, $\KKs \nu_{\tau}$, $\piKspizero \nu_{\tau}$, 
$\KKspizero \nu_{\tau}$, are obtained using this formula. 

Statistical uncertainty is an important issue for the modes with two $\Ks$'s: 
$\tauTO \piKsKs \nu_{\tau}$ and $\tauTO \piKsKspizero \nu_{\tau}$.
For these modes, we use all one-prong decay modes on the tag side and 
determine the branching fraction using the luminosity measured using 
the Bhabha process:
\begin{align}
\mathcal{B}_{i} = \frac{N^{\rm Sig}_{i}}{ 2 N_{\tau\bar{\tau}} \mathcal{B}_{1 - \rm prong}},
\label{eq:bfatag}
\end{align}
where $\Br_{1 - \rm prong}$ is the one-prong decay branching fraction of $(85.35 \pm 0.07)\%$.
$N_{\tau\tau} = \Lum \sigma_{\tau\tau}$ is the number of produced $\tau$ pairs 
determined from the luminosity, $\Lum =(669\pm 9) {\rm fb}^{-1}$, and the 
$\tau$-pair production cross section 
$\sigma_{\tau\tau} = (0.919\pm 0.003) {\rm nb}$~\cite{Banerjee:2007is}.
$N^{\rm Sig}_{i}$ is the number of signal events after efficiency and background 
corrections. 

In both cases, the number of signal events is determined simultaneously 
by using the inverse efficiency matrix to take into account the cross-feed 
from one decay mode into another:
\begin{align}
N^{\rm Sig}_{i} = \sum_{j} (\mathcal{E}^{-1})_{ij}(N^{\rm Data}_{j} - N^{\rm Bg}_{j}),
\label{eq:nsigexp}
\end{align}
where $i$ represents the true decay mode of interest and $j$ represents the 
reconstructed decay mode.
$N^{\rm Data}_{j}$ is the number of selected events in the $j$-th decay mode 
and $N^{\rm Bg}_{j}$ 
is the background coming from decay modes other than the six modes 
under consideration together with the 
non-$\tau$ processes. Hereinafter, we use ``background" to mean this.   
$\mathcal{E}^{-1}$ is the inverse of the selection efficiency matrix
($\mathcal{E}_{ji}$ being the probability of reconstructing a true decay type $i$ 
as a decay type $j$).

\subsection{Background and efficiency \label{sec:backgroundandefficiency}} 

\subsubsection{Background}

The number of background events from  $\tau$ decays other than the six modes
analyzed here is determined by the {\tt TAUOLA} MC using the world-average 
(PDG) branching fractions~\cite{Beringer:1900zz}.
The uncertainties of the PDG branching fractions are used as a 
measure of the  background  uncertainty.

The non-$\tau$ decay contributions are dominated by $q\bar{q}$ 
continuum events. The background from $q\bar{q}$ for each mode is 
confirmed with the data and MC simulation control sample.
The control sample is prepared with the same selection criteria as the signal, 
but requiring that the invariant mass of the hadron system be larger than
the $\tau$ mass. With this selection, one eliminates the $\tau$-pair events 
and enhances the number of $q\bar{q}$ events.
The number of selected events in data and MC are found to be consistent 
within 20\%.
From this calibration, the $q\bar{q}$ background is found to be 0.2--0.8\% 
for the one-$\Ks$ categories.
On the other hand, the two-$\Ks$ categories have large $q\bar{q}$ background:
the fraction is 8--12\%.
The difference between data and MC in the control region is taken as a 
systematic error of the $q\bar{q}$ background estimation.
Backgrounds from $B\bar{B}$ and two-photon processes are negligible: 
0.1--0.5\% for two-photon events and $<$ 0.1\% for $B\bar{B}$.
The fraction of the total background for each mode, summarized in the 
fourth column of Table~\ref{tab:selection}, ranges from 2.4\% to 12\%. 

\subsubsection{Calibration and corrections}
 \label{sec:efficiency}

The particle identification efficiencies, as well as the $\Ks$ and $\pi^0$ 
reconstruction efficiencies, are critical issues for this analysis and 
difficult to reproduce using MC with the required precision; it is necessary to 
calibrate them using data. 
For this purpose, several control samples are prepared for data and MC 
in order to check the reliability of the MC simulation, and correction 
tables are constructed.

The calibration of the particle identification efficiency for charged pions 
and kaons is carried out using kinematically identified 
$D^{*-} \to D^{0} \pi^{-}$($D^{0} \to K^{-} \pi^{+}$) decays, where the kaon 
and pion from the $D^0$ decay are known from the charge of the 
accompanying slow pion.
We evaluate the identification efficiencies and misidentification 
probabilities for this calibration sample and compare them to MC expectations. 
From this comparison, we obtain correction factors as a function of 
track momentum and polar angle and apply these to the MC.
The average correction factor for pions (kaons) is $0.971\pm0.007$ ($1.002 \pm 0.001$).
The accuracy of the correction factor, which is a source of the 
systematic uncertainty for the evaluation of the branching fraction, 
is limited by the statistical uncertainties of the kaon and pion sample 
from $D^{*-}$ decays in certain momentum and angular bins and the 
uncertainty of the $D^{*-}$ signal extraction.

The calibration of the efficiency for electrons and muons is carried 
out using two-photon events from the reaction 
$\epm \to e^+ e^- \ell^+ \ell^-$($\ell = e,\mu$).
The efficiency correction table constructed for the two-dimensional space 
of momentum and polar angle in the laboratory frame
and then applied to the Monte Carlo efficiencies.
In this way, the uncertainty on the lepton efficiency is determined by the 
statistics of the two-photon data sample and its long-term stability. 
The latter is evaluated from the variation of the corrections calculated 
using time-ordered subsets of the experimental two-photon data.
The average corrections are $0.981 \pm 0.008$ for electrons
and $0.958 \pm 0.005$ for muons.

The reconstruction efficiency for the $\Ks$ as a function of momentum 
has been studied by using a control sample from the decay chain 
$D^{*-} \rightarrow D^{0} \pi^{-}, D^{0}\rightarrow \Ks \pi^{+}\pi^{-}$.
The number of $\Ks$ signal events that satisfy the full selection is compared 
with the value determined by only fitting the invariant mass distribution without 
any requirements on the secondary vertex reconstruction.
The average correction factor is 0.979 $\pm$ 0.007.

The $\pi^{0}$ reconstruction efficiency is studied using a sample in which
both $\tau$ leptons decay into $h^\pm \pi^{0} \nu_{\tau}$, 
where $h^\pm$ stands for $\pi^\pm$ or $K^\pm$. 
In the study, we first measure the ratio
\begin{align}
  R_{i} =
       N(h^{-} \pi^{0} \nu_{\tau} | h^{+} \pi^{0} \bar{\nu}_{\tau} )/
        N(h^{-} \pi^{0} \nu_{\tau} | \ell^{+} \nu_{\ell} \bar{\nu}_{\tau}) 
\end{align}
for experimental data and the MC ($i$=data, MC).
Here, $N(h^{-} \pi^{0} \nu_{\tau} | h^{+} \pi^{0} \bar{\nu}_{\tau} )$ is the 
number of events with both $\tau$ leptons decaying to $h^{\pm} \pi^{0} \nu_\tau$ 
($double~h\pi^{0}$), while 
$N(h^{-} \pi^{0} \nu_{\tau} | \ell^{+} \nu_{\ell} \bar{\nu}_{\tau}) $ is the number 
of events where one $\tau$ decays to $h^{-} \pi^{0} \nu_{\tau}$ and the 
other to $\ell^{+} \bar{\nu_{\ell}}\nu_{\tau}$  ($single~h\pi^{0}$).
We then take  the  double ratio  
$R=R_{\rm data}/R_{\rm MC}$ in which many 
common factors, such as  the normalization and tracking efficiency, cancel. 
If we rely on the world-average branching fractions for 
$\tau^{-} \to h^- \pi^{0}\nu_{\tau}$ and 
$\tau^{-} \to \ell^{-} \bar{\nu}_{\ell} \nu_{\tau}$, the double ratio depends 
on the product of the corrections of the $\pi^{0}$ reconstruction and 
lepton ID efficiencies only, where the latter is well-known from the 
two-photon events as well as other studies.
From the double ratio $R$, the MC-data correction for the $\pi^{0}$ efficiency 
is determined to be $R = 0.957 \pm 0.015$. 
A result consistent with this value is also obtained from a study using 
$\eta$ decays, where the ratio of the number of $\eta$ events reconstructed 
from  $\eta \to \gamma\gamma$ and $\eta \to 3\pi^{0}$ is compared in 
experimental data and MC.

\begin{table*}
\caption{Probabilities $\mathcal{E}_{ji}$ of the efficiency matrix for reconstructing a true decay type $i$ as a decay type $j$, in \%, 
for the six decay modes under study.
The first four rows shows the efficiency matrix for lepton tagging, 
while the last two rows show efficiencies for lepton and hadron tagging.
The efficiencies include the $\Ks \to \pi^+\pi^-$ branching fraction.
The dash indicates values smaller than 0.01\%. 
\label{tab:effmatrix}}
\begin{tabular} { l c c c c c c }
\toprule 
Selected  & \multicolumn{6}{c}{ True Decay mode} \\
decay mode     & $\piKs$ \ & $\KKs$ \ & $\piKspizero$ \ & $\KKspizero$ \ & $\piKsKs$ \ & $\piKsKspizero$ \\   \colrule
        $\piKs$             & 7.09 & 1.65 & 1.07 & 0.31 & 0.67 & 0.13 \\
        $\KKs$              & 0.35 & 6.69 & 0.06 & 1.01 & 0.04 & 0.01 \\
        $\piKspizero$       & --- &  ---  & 2.65 & 0.54 & 0.51 & 0.23 \\
        $\KKspizero$        & --- &  ---  & 0.11 & 2.19 & 0.01 & ---  \\
        $\piKsKs$           & --- &  ---  & ---  & ---  & 2.47 & 0.53 \\
        $\piKsKspizero$     & --- &  ---  & ---  & ---  & 0.04 & 0.81 \\
\botrule
\end{tabular}
\end{table*}

\begin{table*}
\caption{ Uncertainties $[\sigma_{\mathcal{E}}]_{ji}$  on the efficiency matrix
(in \%). The dash indicates values smaller than 0.001\%. 
\label{tab:effmatrixerr}}
\begin{tabular} { l c c c c c c }
\toprule 
Selected          & \multicolumn{6}{c}{ True Decay mode} \\
 decay mode     & $\piKs$    & $\KKs$   & $\piKspizero$  & $\KKspizero$  & $\piKsKs$ & $\piKsKspizero$ \\   \colrule
        $\piKs$             & 0.119 & 0.069 & 0.018 & 0.013 & 0.011 & 0.002 \\
        $\KKs$              & 0.011 & 0.116 & 0.002 & 0.018 & 0.001 & ---   \\
        $\piKspizero$       & ---   & ---   & 0.060 & 0.025 & 0.009 & 0.005 \\
        $\KKspizero$        & ---   & ---   & 0.004 & 0.050 & ---   & --- \\
        $\piKsKs$           & ---   & ---   & ---   & ---   & 0.071 & 0.015 \\
        $\piKsKspizero$     & ---   & ---   & ---   & ---   & 0.001 & 0.027 \\
\botrule
\end{tabular}

\end{table*}

\subsubsection{Efficiency matrix}

After taking into account the corrections discussed in the previous 
subsection, the efficiency matrix $\mathcal{E}_{ji}$ is obtained. 
The values of $\mathcal{E}_{ji}$ and their uncertainties are summarized 
in Tables~\ref{tab:effmatrix} and \ref{tab:effmatrixerr}, respectively.
For example, the efficiency for selecting a true $\tauTO \piKs \nu_{\tau}$ decay as a 
$\piKs$ or $\KKs$ candidates is $(7.09 \pm 0.12)\%$ and $(0.35\pm 0.01)\%$, respectively.
The uncertainty of the first efficiency is dominated by the uncertainty 
of the pion and lepton 
identification efficiency (0.8\%) and the $\Ks$ reconstruction efficiency 
(1.4\%), while the uncertainty of the 
second efficiency is dominated by uncertainty of the misidentification 
efficiency from pion to kaon (3\%).
(The detailed discussions of these uncertainties are given in the next 
subsections.)

The efficiency for selecting a true  $\tauTO \piKspizero \nu_{\tau}$ decay as 
a $\piKspizero$ or 
$\piKs$ candidate is $(2.65 \pm 0.06)\%$ and $(1.07 \pm 0.02)\%$, respectively. 
The uncertainty of the first efficiency includes the uncertainties 
for the charged pion, $\Ks$ and $\pi^0$ identification. 
The uncertainty of the $\pi^{0}$ identification is estimated to be 1.5\%.
The same uncertainty is assigned to the decays without the $\pi^{0}$ meson. 

It is worth noting that the migration efficiency for 
the modes without $\pi^{0}$ selected as the modes 
with $\pi^{0}$ is negligible because, as mentioned, 
the spurious $\pi^{0}$ mesons are subtracted using 
the events in the $\pi^{0}$ sideband region.

\subsection{Systematic uncertainties}

\label{sec:systematics}

The sources of systematic uncertainties can be categorized as
those related to detection/reconstruction efficiencies and other items 
such as hadron decay models, background estimation, 
normalization and  event selection such as the $\gamma$ veto.
The  efficiencies have several uncertainties, arising from track finding, 
particle identification, 
$\Ks$ and $\pi^{0}$ reconstruction and the $\pi^{0}$ sideband subtraction.

\subsubsection{Uncertainty of tracking and particle identification}

The uncertainty of the charged track finding efficiency is 0.35\% per 
charged track.
Since the track finding uncertainty partially cancels in 
Eq.~(\ref{eq:bfltag}), 
the net uncertainty is 0.7\% for the modes with one $\Ks$ and 2.1\%
for the modes with two $\Ks$, where the uncertainty for tracking efficiency is 
 added linearly assuming 100\% correlation.

The uncertainties due to  particle identification are estimated from the 
precision of the efficiency calibration procedure.
The uncertainty for the pion and kaon efficiency is found to be 
0.4\% and 0.8\%, respectively. 
The uncertainties for misidentification from pion to kaon and vice versa 
are found to be 3\% for each. 
The uncertainty for electron (muon) identification is 0.8\% (0.5\%).

The efficiency for the $\Ks$ reconstruction is studied using a $\Ks$ 
control sample 
from the $D^{*} \rightarrow \pi_{s} D^{0}$, 
$D^{0}\rightarrow \Ks\pi^{+}\pi^{-}$ decay chain by comparing the  
$\Ks$ yields with and without vertex reconstruction (0.7\%) as well as by 
varying the requirements on $dr$, $z_{dist}$, $\ell_{f}$, and the 
$M(\pi^{+}\pi^{-})$ window (1.2\%).
The net uncertainty from $\Ks$ reconstruction is estimated to be 1.4\%.

\subsubsection{Uncertainty of $\pi^{0}$ reconstruction}

The uncertainty due to the correction of the $\pi^{0}$ efficiency is 
determined by using $\tau^{-} \to h^{-} \pi^{0} \nu_{\tau}$ samples. 
The dominant uncertainty for $\pi^{0}$ efficiency comes from the 
method of counting the number of $\pi^{0}$ events.
Two methods, one using the subtraction of sideband events and the other 
using fits with a logarithmic Gaussian, are used to estimate the signal 
and background $\pi^{0}$.
The uncertainty of the $\pi^{0}$ efficiency is estimated to be 1.5\%.

\begin{table*}
\caption{Summary of the relative statistical and  systematic uncertainties. 
The values in the row ``Efficiency matrix" show the diagonal elements of the 
covariant matrix in the first term of Eq.~(\ref{eq:cov}), 
which correspond to the total uncertainties of the tracking, 
particle identification, and $\pi^{0}$ and $\Ks$ reconstruction efficiencies.
Each contribution is shown as sub-items using parentheses.
The total systematic uncertainty is obtained from the diagonal element of the covariance matrix given in Eq.~(\ref{eq:cov}).
}
\label{tab:systematicseffmatrix} 
\begin{tabular}{ l c c c c c c c } 
\toprule
             & \multicolumn{6}{c}{$\bigtriangleup \Br/\Br$ (\%)}  \\ 
Error Source & $\KsX$ &\ $\piKs$ \ & \ $\KKs$\  & \  $\piKspizero$\ & \ $\KKspizero$\  & \ $\piKsKs$\ & \ $\piKsKspizero$\ \\
\colrule
\ Statistical uncertainty& 0.2 & 0.3 &  0.9&  0.8& 1.3 &  1.4& 10.8    \\ \hline
\ Efficiency matrix      & 1.6 & 1.7 & 2.1 & 2.3 & 2.4 & 2.9 & 4.0 \\
\ \ \ Track finding      & (0.7) & (0.7) & (0.8) & (0.7) & (0.7) & (2.1) & (2.7) \\
\ \ \ Particle ID        & (---)  & (0.6) & (1.0) & (0.7) & (0.8) & (0.8) & (1.0) \\
\ \ \ $\Ks$ reconstruction & (1.4)  & (1.4) & (1.5) & (1.4) & (1.4) & (1.8) & (2.3) \\
\ \ \ $\pi^{0}$ reconstruction &(---) & (0.1) & (0.2) & (1.5) & (1.5) & (0.0) & (1.5) \\
\ Hadron decay model   & --- & --- & 0.7 & 0.3 & 3.4 & 1.2 & 4.2 \\ 
\ Background           & 0.5 & 0.2 & 0.3 & 1.9 & 0.4 & 1.8 & 3.2 \\ 
\ Normalization        & 0.5 & 0.5 & 0.5 & 0.5 & 0.5 & 1.4 & 1.4 \\ 
\ $\gamma$ veto        & --- & 0.1 & 1.8 & 1.2 & 1.5 & 1.0 & 2.0 \\ \colrule
\ Total systematic uncertainty     & 1.7 & 1.8 & 3.7 & 3.5 & 4.9 & 4.0 & 10.1 \\
\botrule
\end{tabular}
\end{table*}

\begin{figure}
\includegraphics[width=0.95\columnwidth]{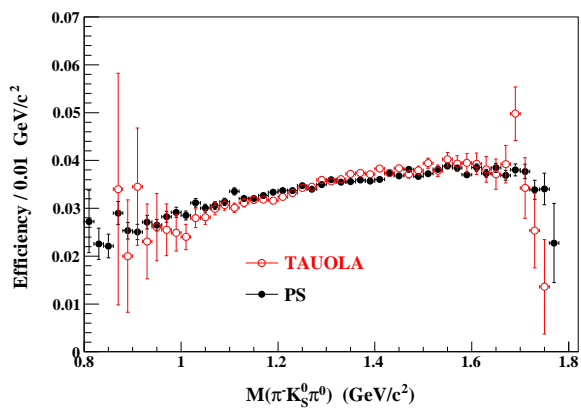}
\caption{(color online) Efficiency as a function of the hadron mass 
for $\tauTO \piKspizero \nu_{\tau}$. 
The open circles are the efficiencies obtained
using the {\tt TAUOLA} event generator.
The closed circles are  obtained assuming a uniform angular decay distributions.
The tagging and branching fraction factors are included in the value 
of the efficiency.}
\label{fig:efficiency}
\end{figure}

\subsubsection{Decay model dependence of the efficiency}

The signal efficiency can potentially change depending on the dynamics of 
the hadronic system. 
A test is performed with a set of MC events generated according to phase space (PS) 
in addition to the standard MC sample based on {\tt TAUOLA}.
For both sets, the invariant mass distribution  for the 
full hadronic system has been tuned to agree with that of experimental data. 
The subsystem mass distribution in the three- or four-body decays 
and their angular distributions differ between the {\tt TAUOLA} and PS models.
The efficiency as a function of the $\piKspizero$ invariant mass 
in $\tau^{-}\to \piKspizero \nu_{\tau}$
is shown in Fig.~\ref{fig:efficiency} for these models.
In both cases, the efficiency changes smoothly as a function of 
hadronic mass and the efficiencies at the same hadronic mass 
agree in both cases except for the mass region above 1.7 GeV/$c^{2}$.
This agreement indicates that the efficiency is insensitive to the 
detailed decay models of the hadronic system.
We obtain the net efficiency for the full mass region in both models 
and assign the difference between them as a systematic uncertainty 
due to the decay model. 
The resultant model dependences   for $\pi\Ks \nu_{\tau}$, $\KKs \nu_{\tau}$, 
$\KKspizero \nu_{\tau}$, $\piKsKs \nu_{\tau}$ and $\piKsKspizero \nu_{\tau}$, 
range from 0.3 to 4.2\% 
as shown in the row labeled 
``Hadron decay model" in Table~\ref{tab:systematicseffmatrix}.

\subsubsection{Uncertainty of the background}

The uncertainty due to the background from other $\tau$ decays 
is estimated from the uncertainties of the world-average branching fractions 
given in the PDG listing~\cite{Beringer:1900zz}. 
The uncertainty of the continuum background is estimated from the 
difference between MC and data for the control sample above the $\tau$ mass.
Adding the uncertainty from other $\tau$ decays and the uncertainty of 
the $q\bar{q}$ continuum in quadrature, the background uncertainties 
for each decay mode are in the range from 0.2\% to 3.2\% 
as shown in Table~\ref{tab:systematicseffmatrix}.

\subsubsection{Uncertainty of the normalization}

The uncertainty due to the normalization is 0.5\% for the modes that 
use $e-\mu$ events for the normalization,
while the uncertainty for $\piKsKs\nu_{\tau} $ and $\piKsKspizero \nu_{\tau}$ is 1.4\%.
The former uncertainty includes the uncertainty of 
$\mathcal{B}(\tauTO l^{-} \nu_{l} \nu_{\tau})$ ($0.1\%$) and
the background uncertainty in $e-\mu$ event selection (less than $0.1\%$).
The latter is dominated by the uncertainty of the luminosity measurement.

\subsubsection{Uncertainty of the $\gamma$ veto}

The uncertainty due to the $\gamma$ veto is obtained by varying 
the condition on the energy sum of extra photons 
$E^{\rm ex}_{\gamma}$ from 0.2 GeV to 1.0 GeV.
The uncertainties for each mode range from 0.1\% to 2.0\% 
as shown in Table~\ref{tab:systematicseffmatrix}.

\begin{table*}
\caption{ Correlation coefficients between the branching fraction measurements.
Both statistical and systematic errors are included.}
\label{tab:totalcovariancematrix}
\begin{tabular}{ l  c c c c c c } \toprule 
                          & $\piKs$  & $\KKs$  & $\piKspizero$ & $\KKspizero$  & $\piKsKs$ & $\piKsKspizero$ \\ \colrule 
$\piKs$      &    1   & -0.230 & -0.132   &  0.023    & -0.019   &  0.004 \\
$\KKs$       &        &   1    &  0.043   & -0.215    & -0.001   &  0.000  \\ 
$\piKspizero$&        &        &     1    & -0.204    & -0.063   &  0.006 \\ 
$\KKspizero$ &        &        &          &  1        &  0.002   &  0.000 \\
$\piKsKs$    &        &        &          &           &   1      & -0.230 \\
$\piKsKspizero$ &     &        &          &           &          &     1     \\ \botrule
\end{tabular}
\end{table*}

\subsubsection{Covariance matrix and error propagation}

Taking into account all uncertainties discussed in the previous sections, 
we obtain the covariance matrix for the
measured branching fractions.
Since the branching fractions are determined simultaneously by 
solving linear equations,
there is a correlation among the results.  
These correlations are taken into account by the covariance matrix.
The full covariance matrix cov($\Br_{i},\Br_{j}$) is given by the formula 
provided in Ref.~\cite{Lefebvre:1999yu},

\begin{align}
{\rm cov}(\Br_{i},\Br_{j}) = f_{a}f_{b} {\rm cov}(\mathcal{E}^{-1}_{ia}, \mathcal{E}^{-1}_{jb})   
    + \mathcal{E}^{-1}_{ik} \mathcal{E}^{-1}_{jl} {\rm cov}(f_{k}, f_{l} ),
\label{eq:cov}  
\end{align}
where the indices indicate the decay modes of interest, and the summation is assumed implicitly 
if the same index is repeated. 
The quantity $f_{j}$ is defined by
$\mathcal{B}_{i} = \sum_{j} \mathcal{E}^{-1}_{ij} f_{j}$ and is given by
\begin{align}
f_{j} =  \frac{ (N^{\rm Data}_{j} - N^{\rm Bg}_{j})}
  {N^{\rm Sig}_{e \textrm{-}\mu}} \frac{\mathcal{B}_{e} \mathcal{B}_{\mu}} {\mathcal{B}_{e}+\mathcal{B}_{\mu}}
\label{eq:bfj3}
\end{align}
and
\begin{align}
f_{j} =  \frac{ (N^{\rm Data}_{j} - N^{\rm Bg}_{j})}
{ 2 N_{\tau\bar{\tau}} \mathcal{B}_{1 - \rm prong}}
\label{eq:bfj4}
\end{align}
for the one $\Ks$ and $\Ks\Ks$ cases, respectively.

The first term in Eq.~(\ref{eq:cov}) represents the covariance 
due to the inverse of the efficiency matrix  $\mathcal{E}_{ji}$.
Assuming that the elements $\mathcal{E}_{ji}$ are uncorrelated,  
the term $ {\rm cov}(\mathcal{E}^{-1}_{\alpha\beta}, \mathcal{E}^{-1}_{ab})$ 
can be expressed as
\begin{align}
{\rm cov}(\mathcal{E}^{-1}_{\alpha\beta}, \mathcal{E}^{-1}_{ab}) =
  (\mathcal{E}^{-1}_{\alpha j} \mathcal{E}^{-1}_{a j})
  [\sigma_{\mathcal{E}}]^{2}_{ji} 
  (\mathcal{E}^{-1}_{i \beta} \mathcal{E}^{-1}_{i b}).
\label{eq:covinveff}  
\end{align}
where $[\sigma_{\mathcal{E}}]_{ji}$ is  the error of ${\mathcal{E}}_{ji}$.
The values of $[\sigma_{\mathcal{E}}]_{ji}$ are summarized in 
Table~\ref{tab:effmatrixerr}.
The error includes the uncertainties due to the track finding, 
particle identification, 
$\pi^{0}$ and $\Ks$ reconstruction efficiencies.

Using Eq.~(\ref{eq:covinveff}), the correlations of the uncertainty for the
track finding, particle identification and $\Ks$ and $\pi^{0}$ reconstruction
efficiencies for the individual modes as well as the cross-feed among
the modes are taken into account.
The total uncertainty as well as each contribution are summarized 
in the row of ``Efficiency matrix'' and its sub-items
in Table~\ref{tab:systematicseffmatrix}.

The second term in Eq.~(\ref{eq:cov}) includes the uncertainties 
from the quantities contained in Eq.~(\ref{eq:bfj3}) and Eq.~(\ref{eq:bfj4}), 
such as the common normalization, the background, and the statistical uncertainty. 
We also include the model dependence and the $\gamma$ veto in this term.

Adding all systematic errors in Eq.~(\ref{eq:cov}),
the total covariance matrix ${\rm cov}(\Br_{i},\Br_{j})$ of the systematic uncertainty is obtained.
The square root of the diagonal element, $\sqrt{{\rm cov}(\Br_{i},\Br_{i})}$,
is given in the last row of Table~\ref{tab:systematicseffmatrix}.
The correlation coefficients, defined as 
${\rm cov}(\Br_{i},\Br_{j})/\sqrt{{\rm cov}(\Br_{i},\Br_{i})~{\rm cov}(\Br_{j},\Br_{j})}$,
are presented in Table~\ref{tab:totalcovariancematrix}, 
where both systematic and statistical 
uncertainties are included. The largest correlation of about $-0.23$ is 
observed for the modes where a charged pion and kaon are interchanged.

\begin{table*}
\caption{Summary of the branching fractions of the $\tau$ lepton decays 
to one or more $\Ks$ obtained in this experiment and previous 
experiments. The first uncertainty is statistical and the second is systematic.
}

\label{tab:results}
\begin{tabular}{c  c c} \toprule
  Mode &   Branching Fraction  & Ref. \\ \colrule
$\KsX \neutau$       & $(9.15\pm 0.01 \pm 0.15)\times 10^{-3}$ &This exp.\\
$\piKs \neutau$      & $(4.16\pm 0.01 \pm 0.08)\times 10^{-3}$ &This exp.\\
$\KKs \neutau$       & $(7.40 \pm 0.07 \pm 0.27)\times 10^{-4}$ & This exp.\\
$\piKspizero \neutau$ & $(1.93\pm 0.02 \pm 0.07)\times 10^{-3}$ &This exp.\\
$\KKspizero \neutau$  & $(7.48\pm 0.10 \pm 0.37)\times 10^{-4}$ & This exp.\\
$\piKsKs \neutau$     & $(2.33 \pm 0.03 \pm 0.09)\times 10^{-4}$ & This exp.\\
$\piKsKspizero \neutau$  & $(2.00\pm 0.22 \pm 0.20)\times 10^{-5}$ &This exp.\\ 
\colrule
$\piKsKl \neutau$  & $(1.01\pm 0.23 \pm 0.13)\times 10^{-3}$ & ALEPH~\cite{Barate:1997tt}\\
$\piKspizeropizero \neutau$  & $(0.13\pm 0.12 \pm 0.00)\times 10^{-3}$ & ALEPH~\cite{Barate:1999hj}\\
$\piKseta \neutau$  & $(0.44\pm 0.07 \pm 0.03)\times 10^{-3}$ &Belle~\cite{Inami:2008ar}\\
$\piKsKlpizero \neutau$  & $(0.31\pm 0.11 \pm 0.05)\times 10^{-3}$ & ALEPH~\cite{Barate:1999hj}\\
$\Kshhh \neutau$  & $(0.115\pm 0.095 \pm 0.04)\times 10^{-3}$ & ALEPH~\cite{Barate:1997tt}\\
\botrule

\end{tabular}
\end{table*}

\begin{figure*}
\includegraphics[width=0.66\columnwidth]{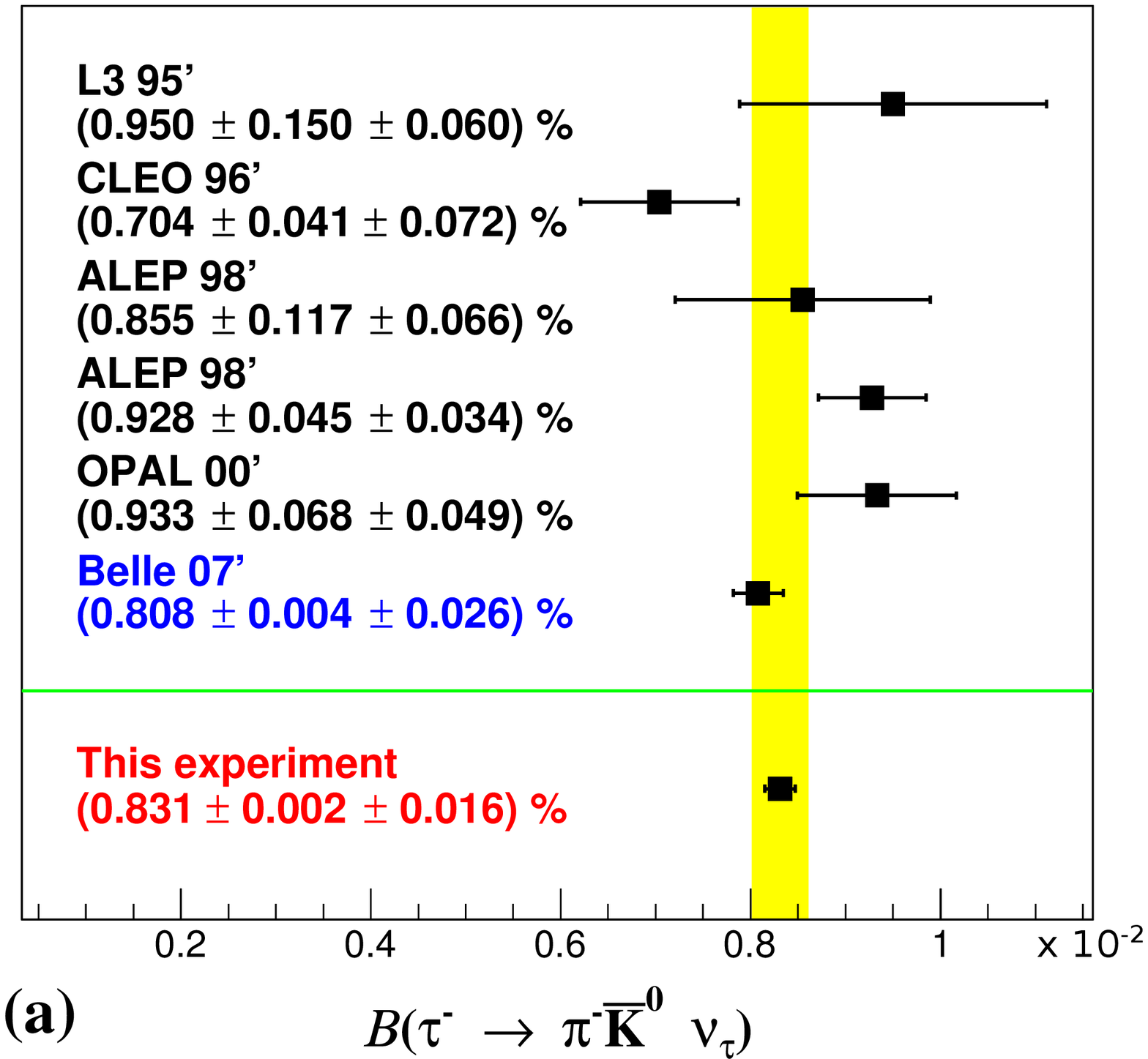}
\includegraphics[width=0.66\columnwidth]{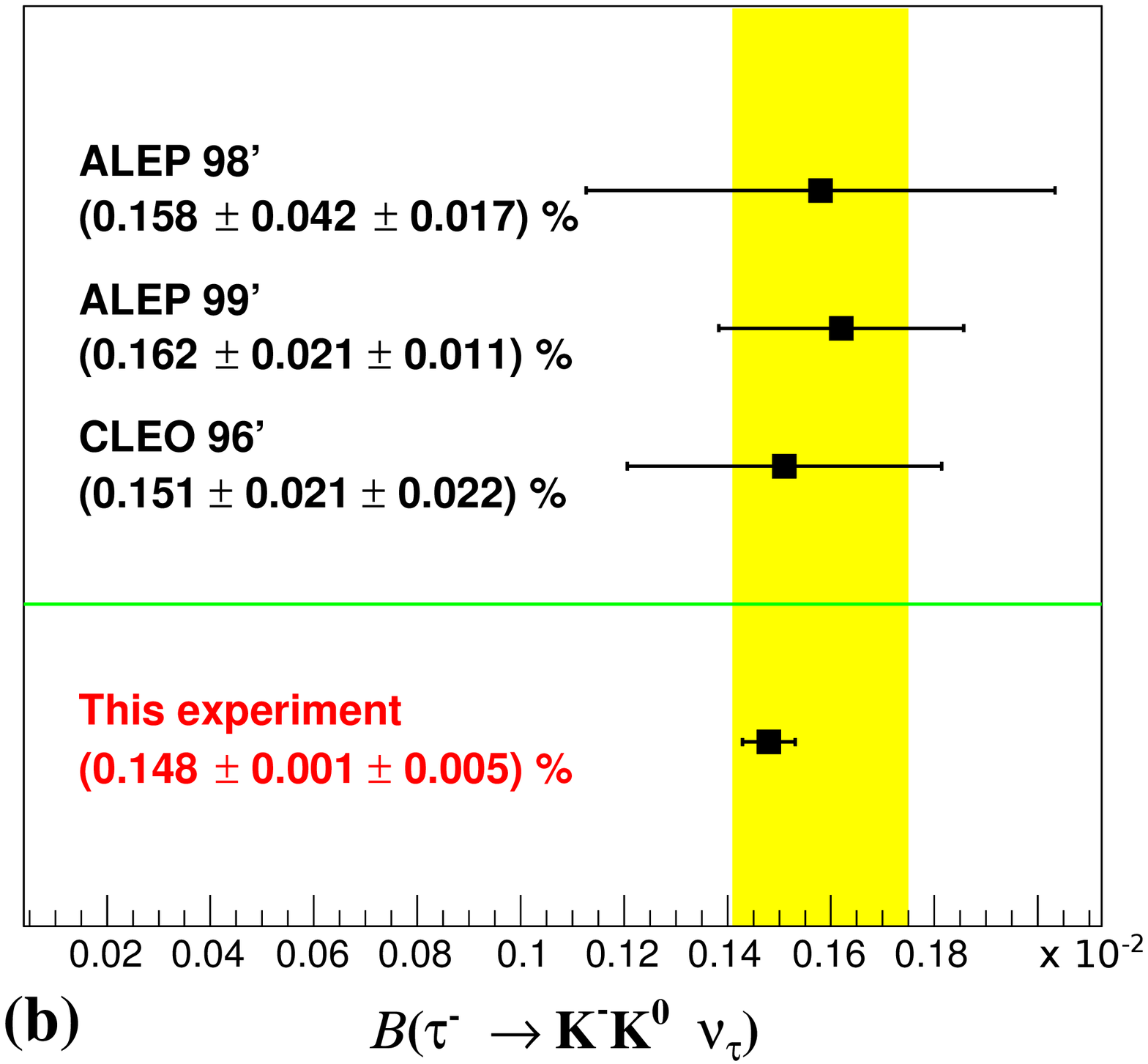}
\includegraphics[width=0.66\columnwidth]{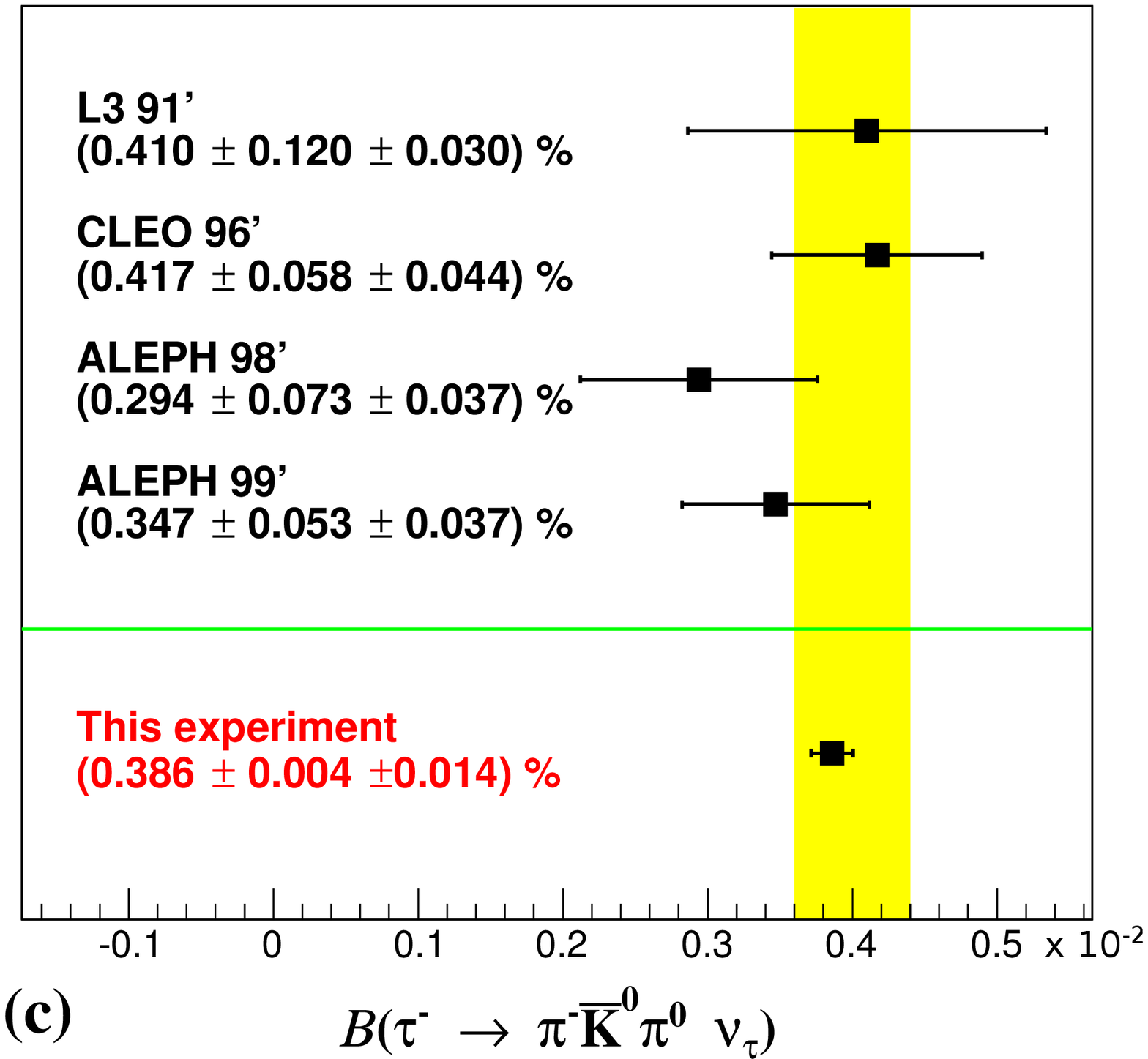}
\includegraphics[width=0.66\columnwidth]{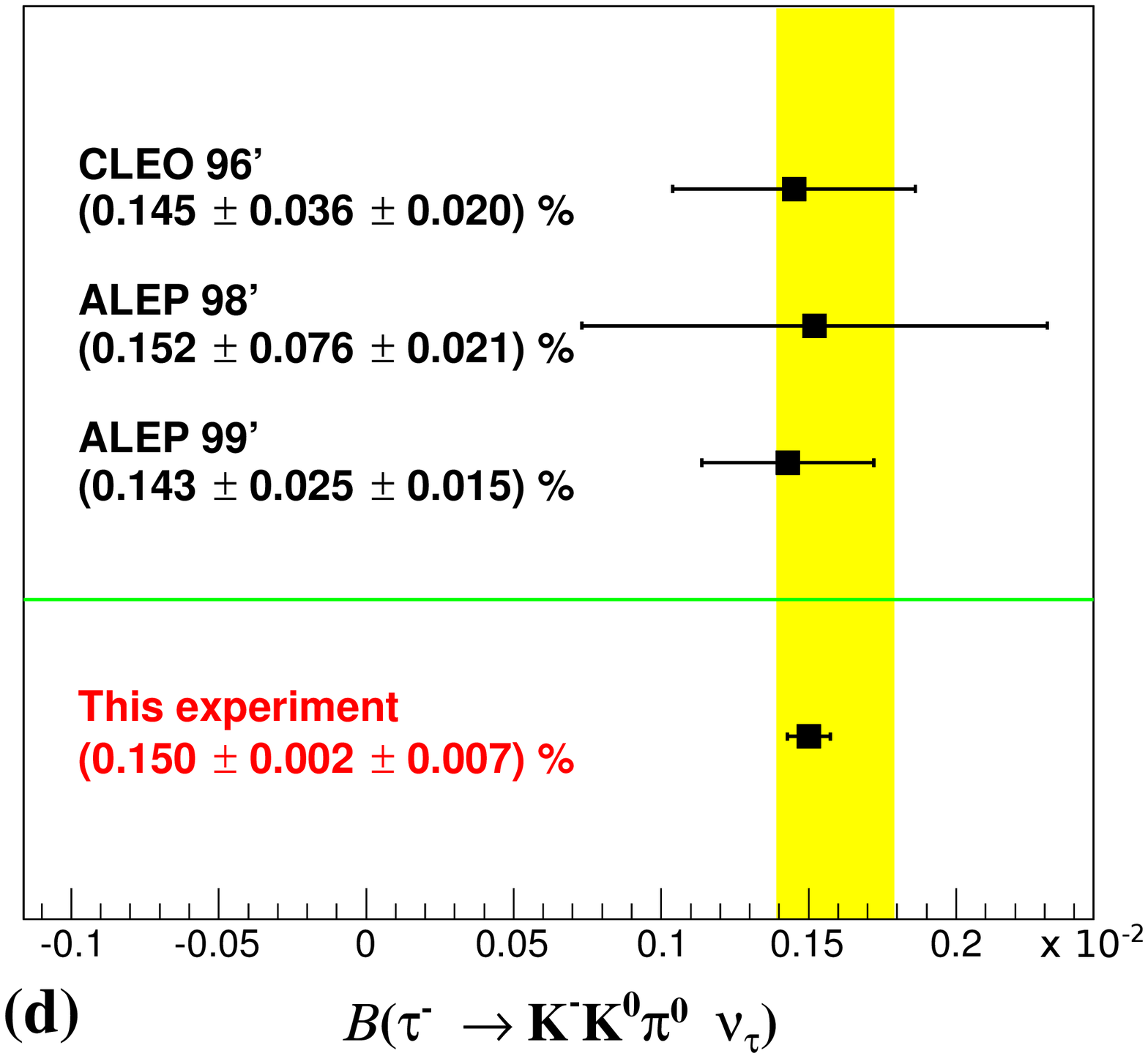}
\includegraphics[width=0.66\columnwidth]{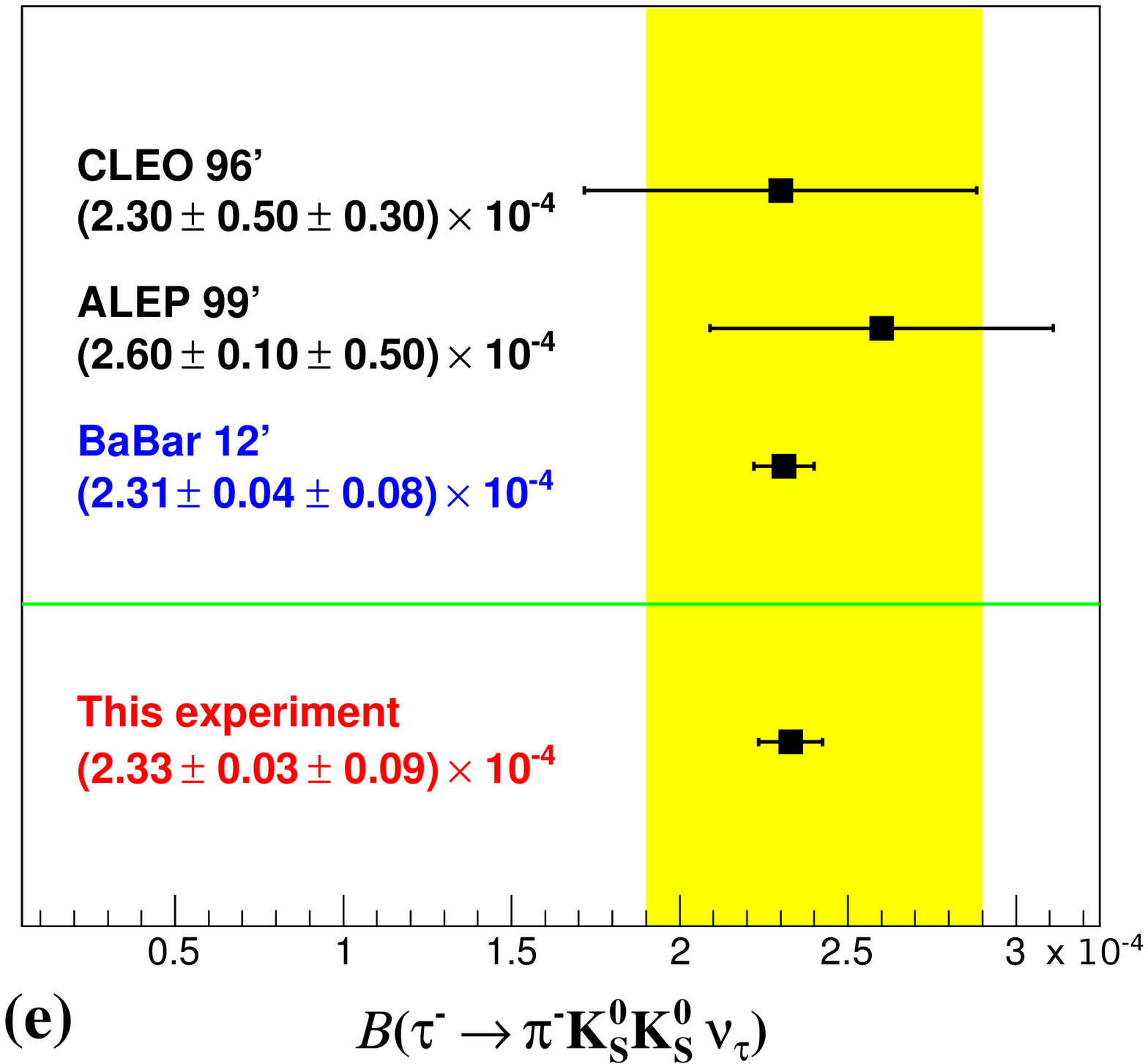}
\includegraphics[width=0.66\columnwidth]{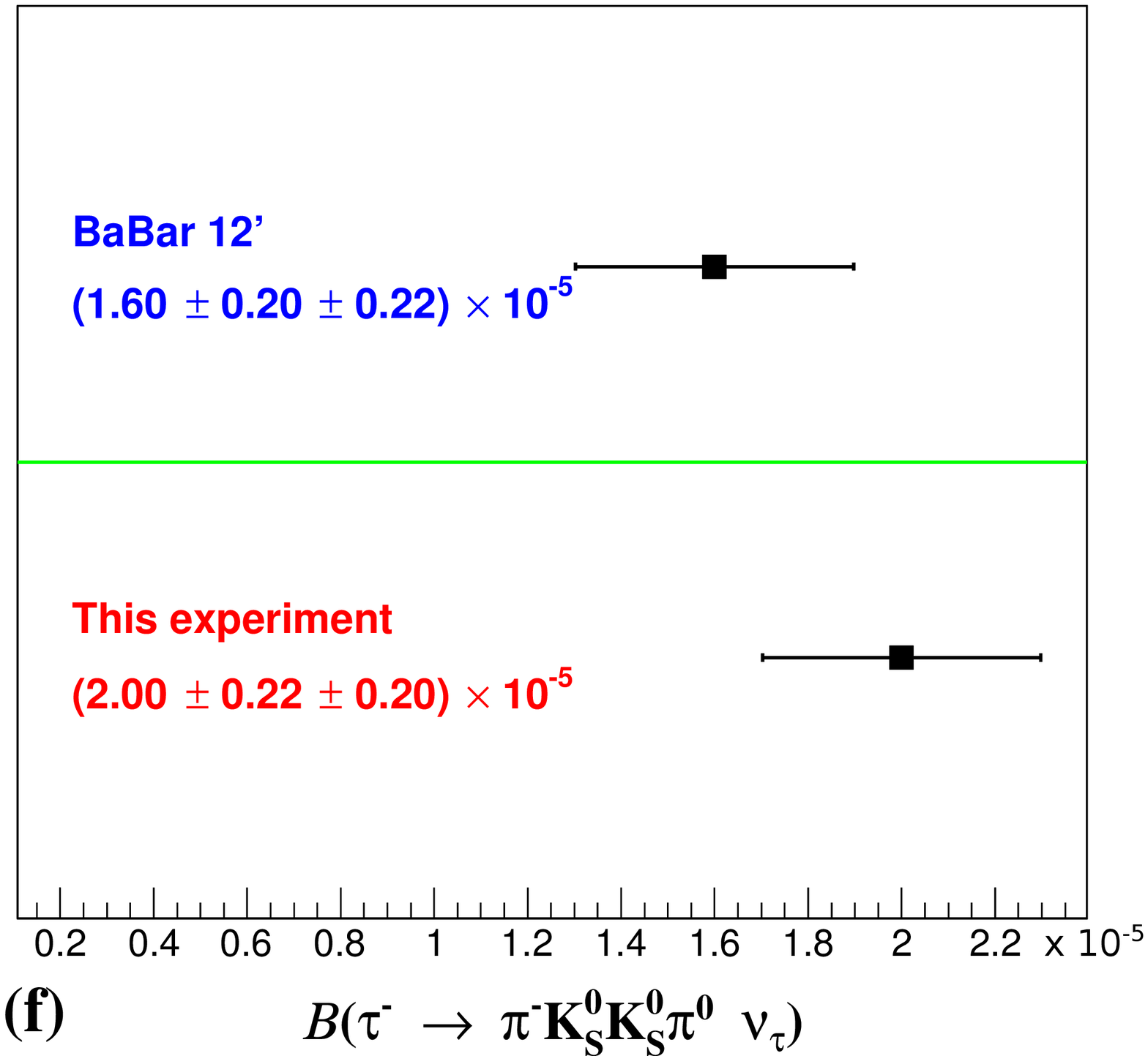}
\caption{(color online) Comparison of results on the branching fractions 
from this work and previous measurements for the six decay modes :
(a) $\pi^{-}\bar{K}^{0}\neutau$, (b) $K^{-}K^{0}\neutau$, 
(c) $\pi^{-}\bar{K}^{0}\pi^{0}\neutau$, (d) $K^{-}K^{0}\pi^{0}\neutau$, 
(e) $\piKsKs\neutau$ and (f) $\piKsKspizero\neutau$.
The band represents the pre-$B$-factory world averages and their uncertainties~\cite{Beringer:1900zz}.}
\label{fig:bfs}
\end{figure*}

\subsection{Branching fractions and discussion}

\subsubsection{Inclusive branching fraction}

The branching fraction for inclusive $\Ks$, 
$\mathcal{B}(\tau^-\to K_{S}^{0} X^- \nu_{\tau})$, 
is determined from the total size of the inclusive $\Ks$ sample 
discussed in Section \ref{subsec:selectionofeventscontainingoneks} 
using Eq.~(\ref{eq:bfltag}).
By applying the corrections for the PID and $\Ks$ reconstruction, 
the signal efficiency is $(9.66 \pm 0.15)\%$ 
while the background admixture is $(4.20 \pm 0.17)\%$ 
among the total selected events.
The background is dominated by the $q\bar{q}$ continuum.
The systematic uncertainty is estimated to be 1.7\%. 
The resulting branching fraction is 
\begin{align}
\mathcal{B}(\tauTO \Ks X^{-} \nu_{\tau}) = (9.15 \pm 0.01 \pm 0.15) \times 10^{-3}.\notag
\end{align}

\subsubsection{Exclusive branching fractions}

The branching fractions of the six exclusive modes, $\piKs \neutau$, 
$\KKs \neutau$, $\piKspizero \neutau$,
$\KKspizero \neutau$, $\piKsKs \neutau$ and $\piKsKspizero \neutau$,
are summarized in Table~\ref{tab:results}.
The precision ranges from 1.8\% to 7.5\% and the systematic uncertainty 
is dominant except for the mode $\tau^{-} \to \piKsKspizero \neutau$.

Figure~\ref{fig:bfs} compares the branching fractions 
obtained in this  and previous experiments.
Assuming that $K^{0}-\bar{K}^{0}$ mixing is negligible, the branching fractions 
involving $K^{0}$ are twice those with $\Ks$.
The accuracy of the branching fractions is improved by a factor of five 
to ten compared to the pre-$B$-factory experiments.
The branching fraction for $\tauTO \piKs \nu_{\tau}$ is consistent with 
our previous result~\cite{Epifanov:2007rf} with improved precision
and supersedes our previous result.
Our result also agrees with BaBar ($\Br(\tau^{-}\to\pi^{-}\bar{K}^{0} \neutau) = 
(8.40 \pm 0.03 \pm 0.23) \times 10^{-3}$~\cite{Aubert:2008an}) within uncertainties.
Recently, the branching fraction for $\tau^{-}\to \pi^{-} \bar{K}^0 \nu_{\tau}$
has been estimated using the crossed channel branching fraction
$\Br(K \to \pi e \bar{\nu}_{e})$ and the measured $K_{S}^{0}\pi^-$ mass spectrum~\cite{Antonelli:2013usa}.
The result is $\Br(\tau^{-}\to \pi^{-} \bar{K}^0 \nu_{\tau})_{\rm Kaon}  = (8.57 \pm  0.30) \times 10^{-3}$.
Our result is consistent with this prediction within uncertainties.

The branching fractions for 
$\tau^{-} \to \KKs \neutau$ and $\tau^{-} \to \KKspizero \neutau$ are 
measured for the first time at the $B$-factories.
The results are consistent with the previous experiments and have 
better precision.
For $\tau^{-} \to \piKspizero \neutau$, the branching fraction 
is measured at the 4\% level by Belle and BaBar, with a marginal 2.5$\sigma$
difference between two experiments.
Recently, BaBar has reported the branching fractions for the $\piKsKs \neutau$ 
and $\piKsKspizero \neutau$ modes~\cite{Lees:2012de}.
Our results agree with those of BaBar within errors.

The sum of all exclusive branching fractions with $\Ks$'s measured in 
this experiment is $(7.83\pm 0.12)\times 10^{-3}$. By adding the branching 
fractions of other modes containing one or more $\Ks$'s but not measured 
in this experiment (see Table~\ref{tab:results}), we obtain the total sum 
of $(9.39\pm 0.35)\times 10^{-3}$, in agreement with the inclusive result of 
$(9.15\pm 0.01\pm 0.15)\times 10 ^{-3}$ within errors. The precision 
of the exclusive sum is dominated by the uncertainties of the branching 
fractions of the modes containing $K_{L}^{0}$.

\begin{figure*}
\includegraphics[width=\columnwidth]{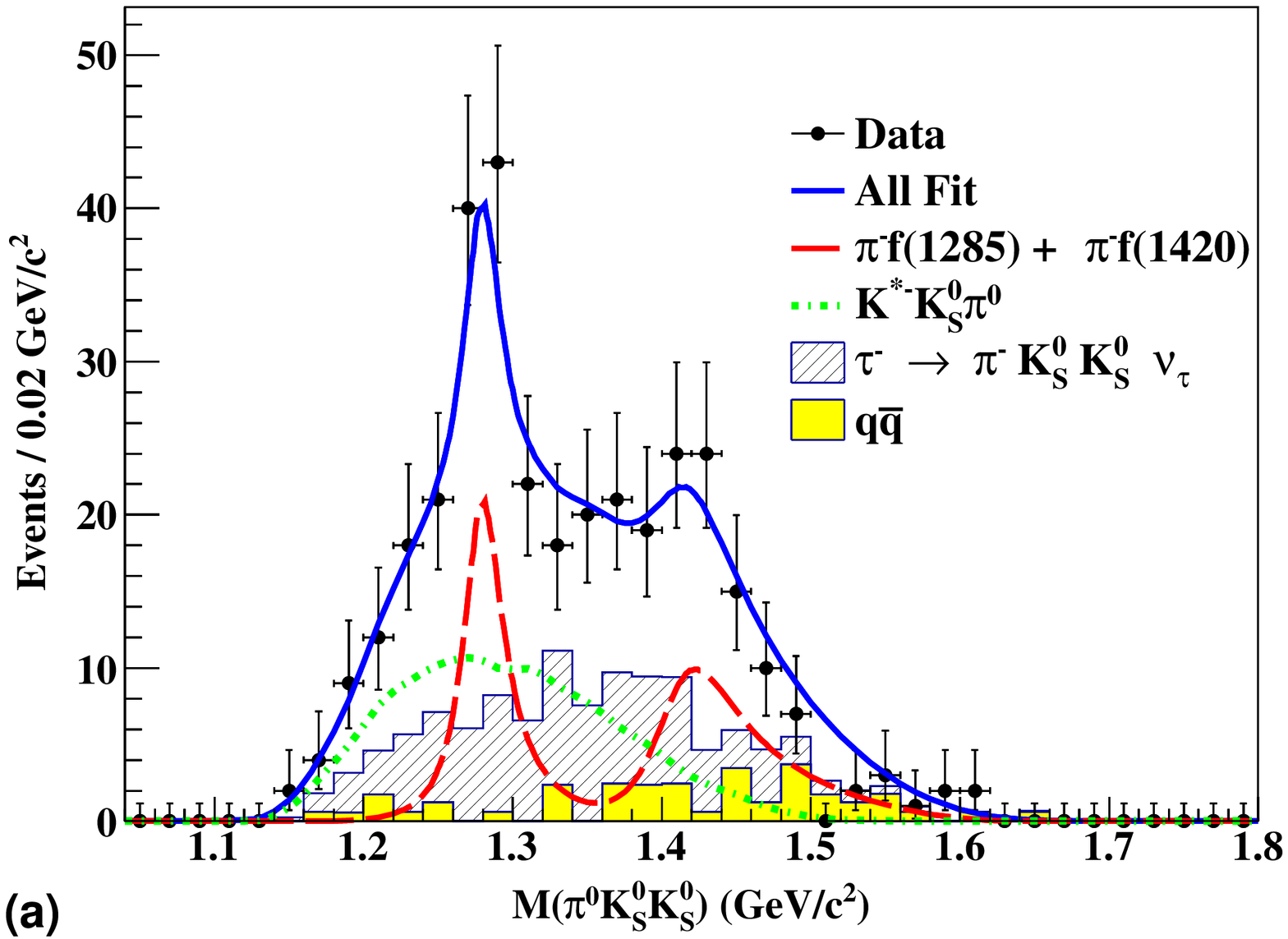}
\includegraphics[width=\columnwidth]{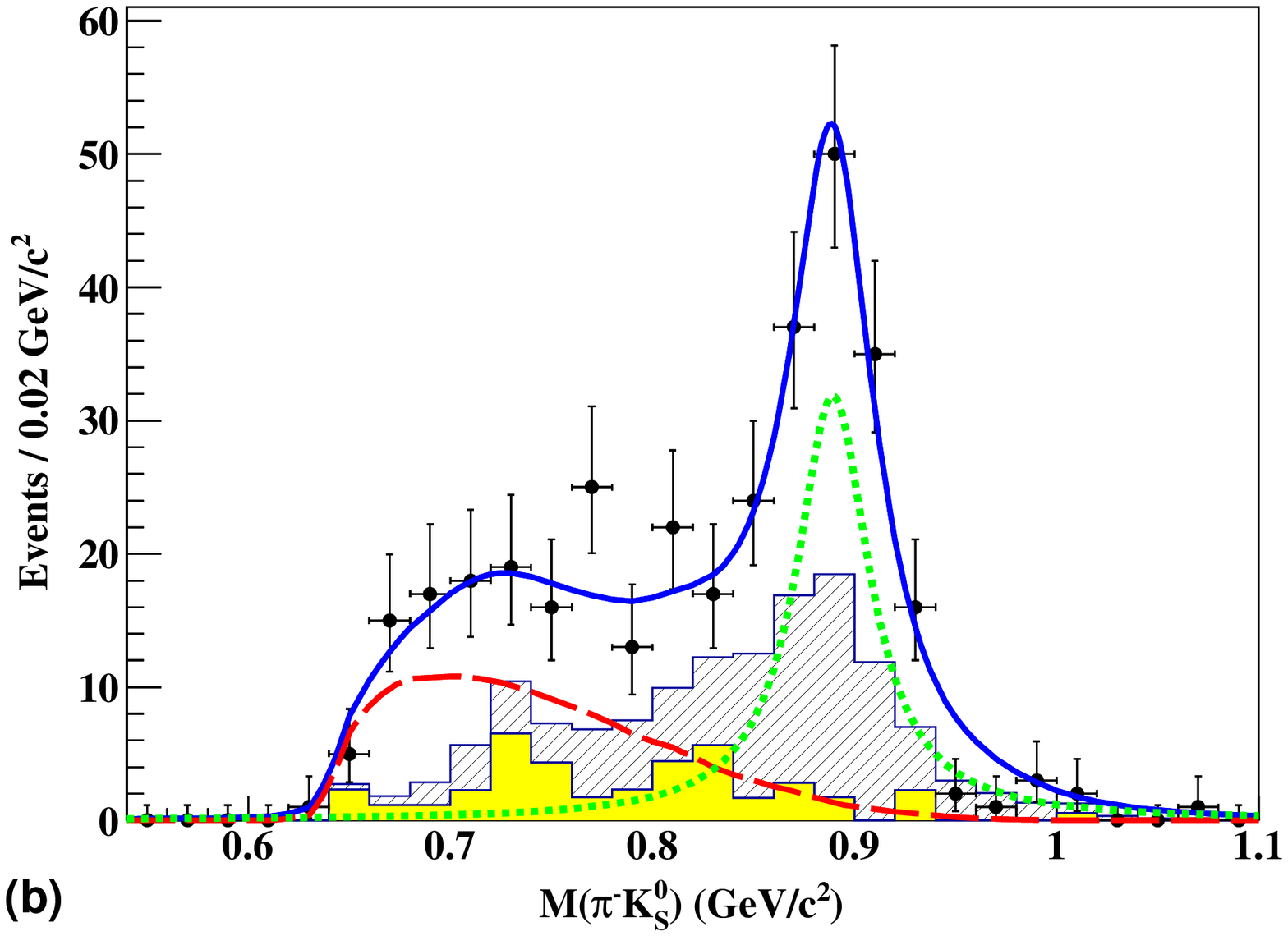}
\includegraphics[width=0.65\columnwidth]{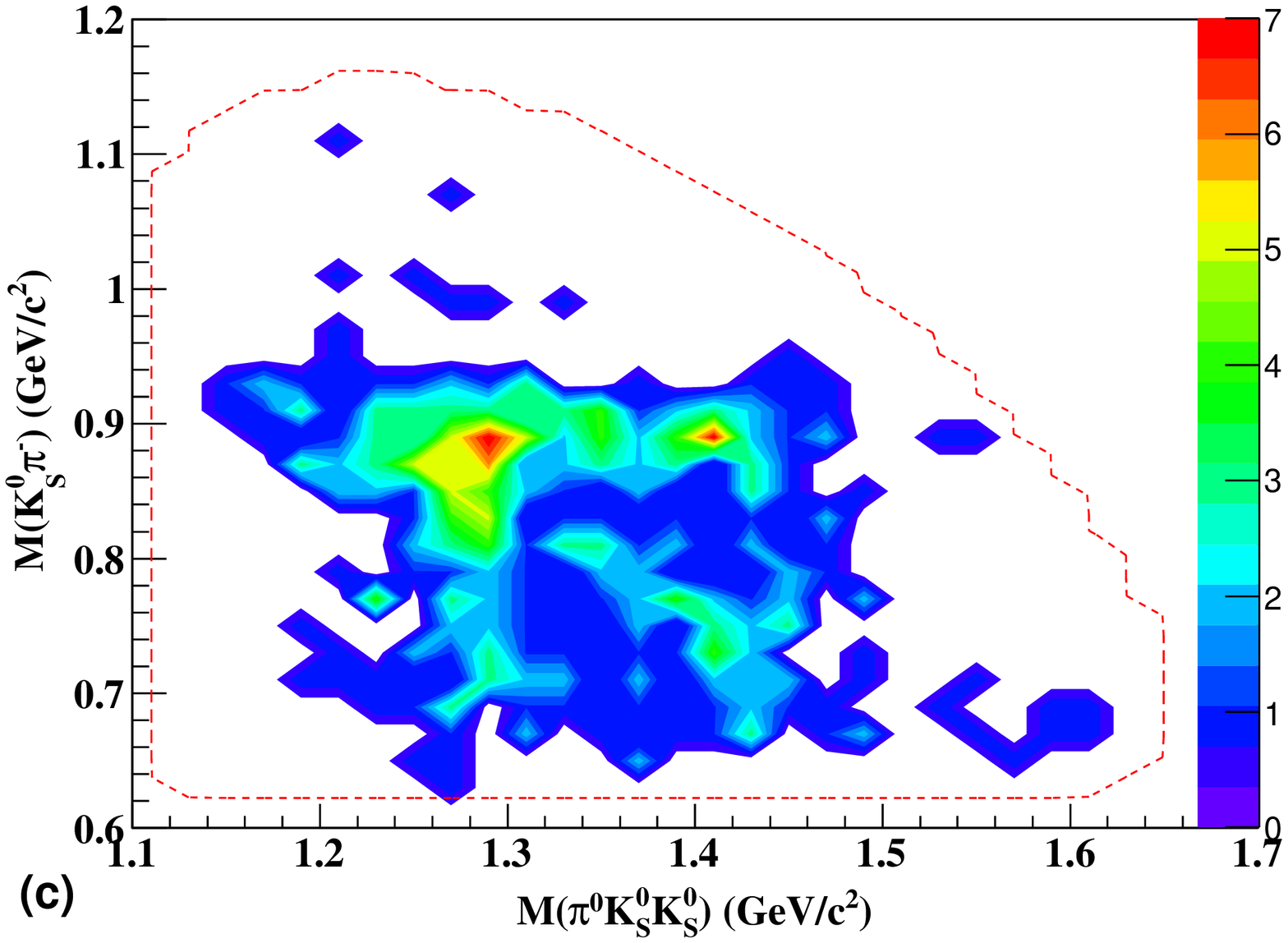}
\includegraphics[width=0.65\columnwidth]{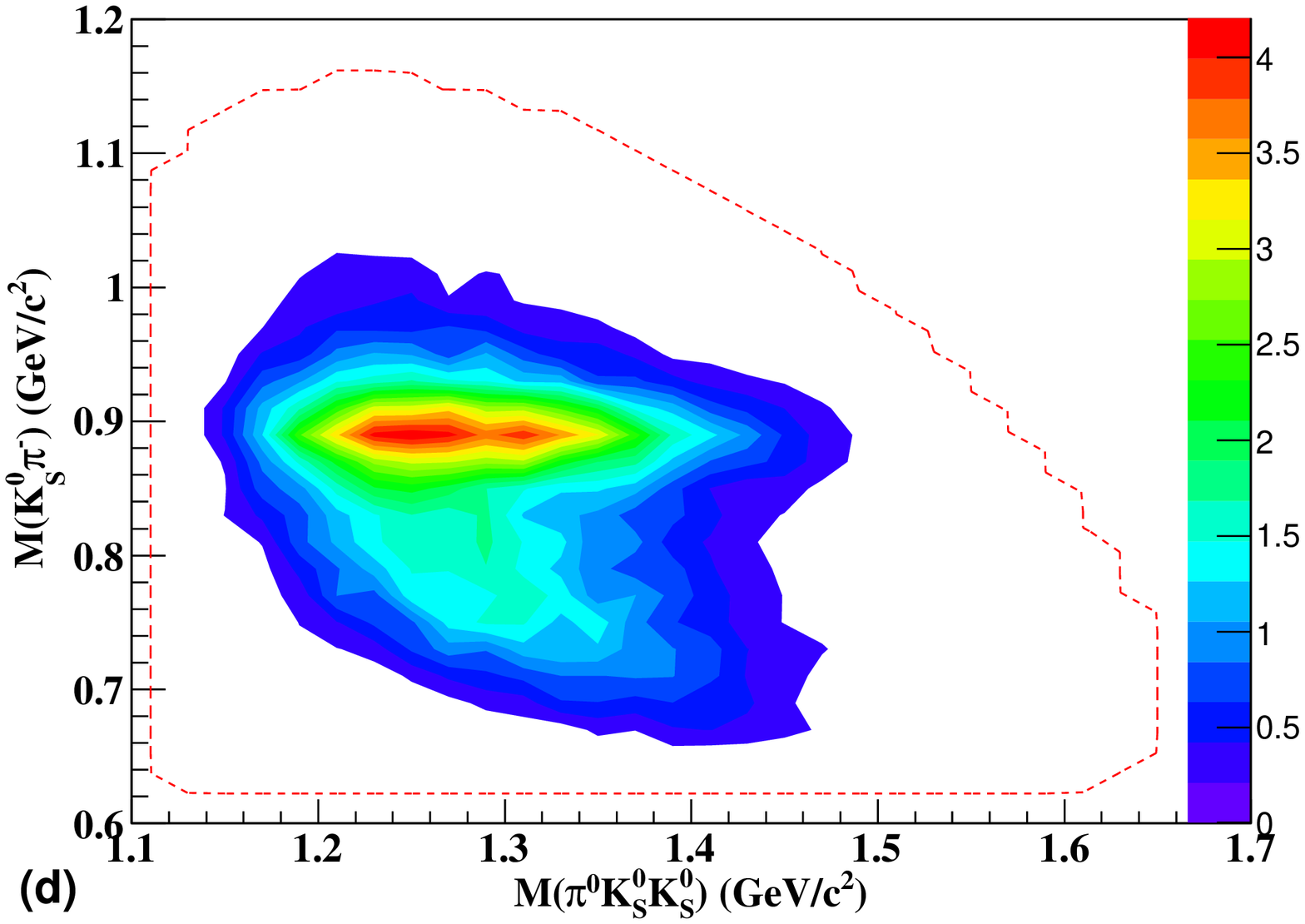}
\includegraphics[width=0.65\columnwidth]{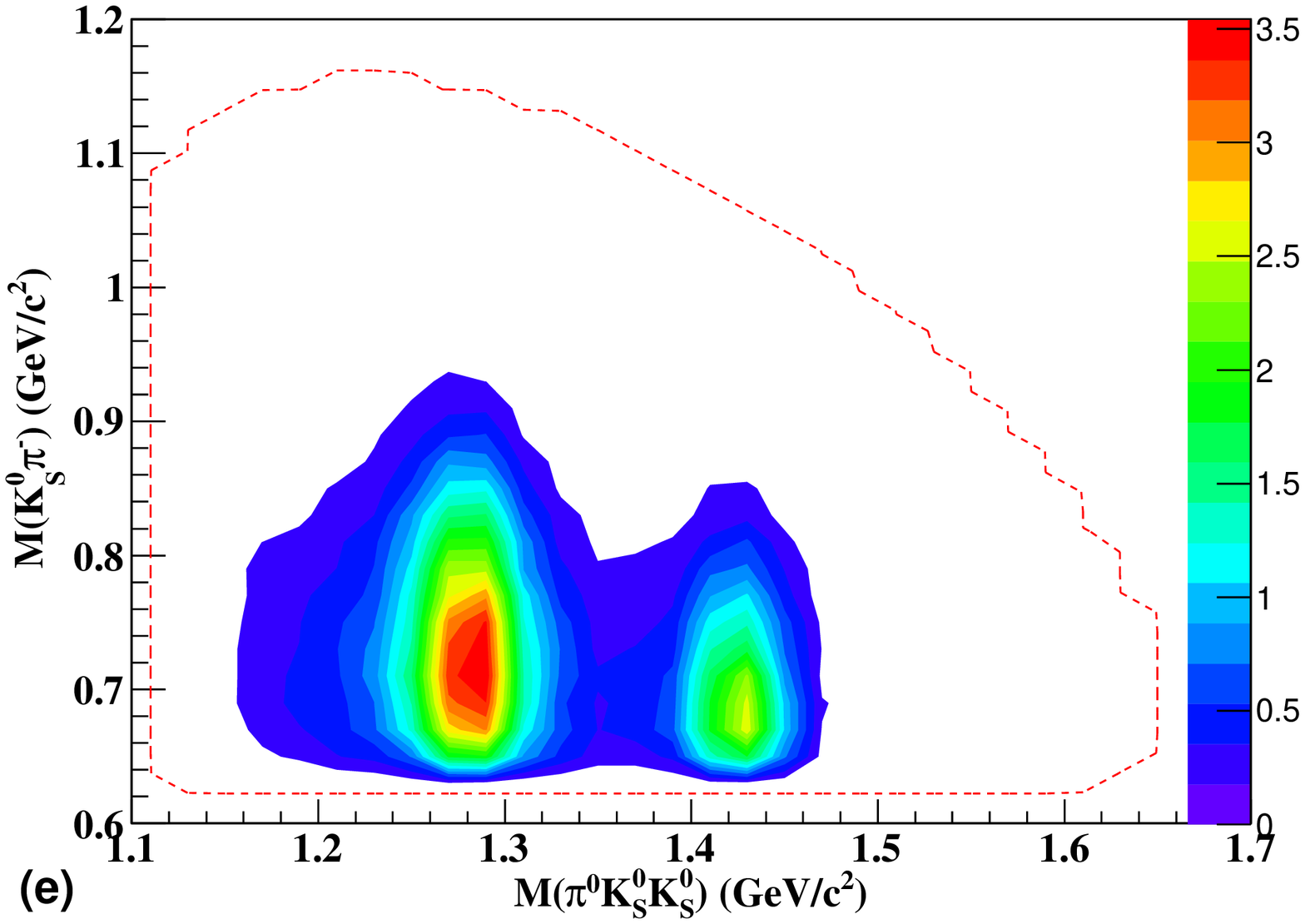}
\caption{
(color online)  Invariant mass of the (a) $\pi^{0}\Ks\Ks$ and (b) 
$\Ks\pi^{-}$ subsystem for 
$\tauTO \piKsKspizero \nu_{\tau}$ candidates.
In both histograms, the solid circles with error bars are data,
the hatched histogram is the background from $\tau^{-} \to \piKsKs \nu_{\tau}$, and 
the shaded (yellow) histogram is the $q\bar{q}$ background.
The solid line is the result of the fit with the 
$\tau^-\to \fone(1285)\pi^- \nu_{\tau}$ +$\tau^-\to \fone(1420)\pi^- \nu_{\tau}$, 
 $\tau^-\to K^{*-}\Ks \nu_{\tau}$ and background contributions.
The $\fone(1285)\pi^- \nu_{\tau}$ +$\fone(1420)\pi^- \nu_{\tau}$ and 
$K^{*-}\Ks \nu_{\tau}$ contributions are shown by the dashed (red) and 
dotted (green) line, respectively.
The two-dimensional plot of the invariant masses of $\Ks\pi^{-}$ and 
$\pi^{0}\Ks\Ks$ system for (c) $\tauTO \piKsKspizero \nu_{\tau}$ candidates in data.
Two-dimensional plots of the MC events for (d) $\tau^{-} \to K^{*-}\Ks\pi^{0}\nu_{\tau}$ and
(e) $\tau^{-} \to \pi^{-}\fone(1285)\nu_{\tau} + \pi^{-}\fone(1420) \nu_{\tau}$ processes.
The dotted curve in (c)-(e) shows the kinematic boundary where the invariant mass of 
the $\piKsKspizero$ system is equal to the $\tau$-lepton mass.
}
\label{fig:submass}
\end{figure*}

\section{Mass spectra in the  $\tauTO \piKsKspizero \nu_{\tau}$ sample}

\begin{figure*}
\includegraphics[width=\textwidth]{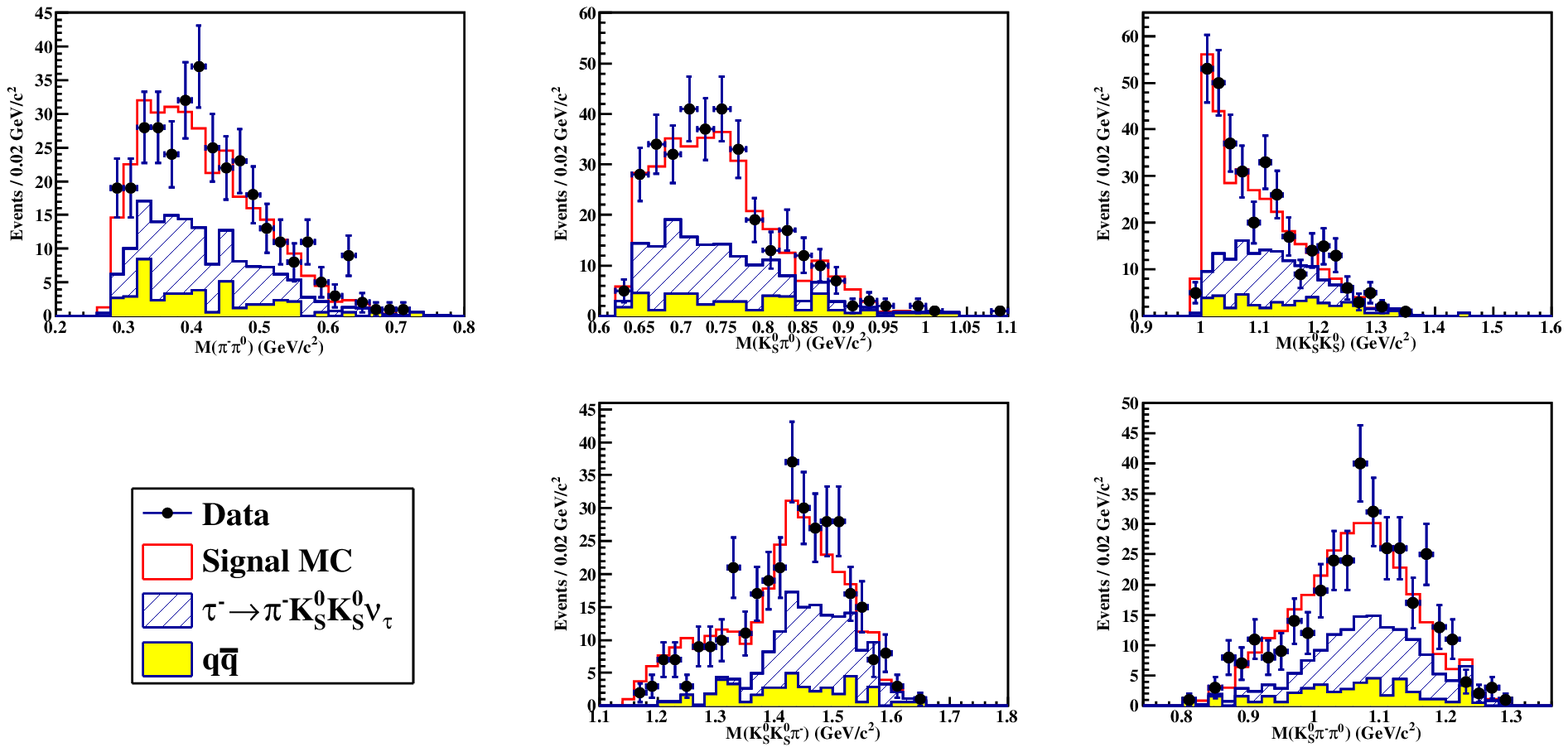}
\caption{(color online) 
Invariant mass distributions of the sub-mass systems for $\tau^{-}\to\piKsKspizero\nu_{\tau}$;
$M(\pi^{-}\pi^{0})$, $M(\Ks\pi^{0})$, $M(\Ks\Ks)$, $M(\Ks\Ks\pi^{-})$ and $M(\Ks\pi^{-}\pi^{0})$. 
The solid circles with error bars are data. 
The blank (red) histogram is the sum of the signal $\tauTO \piKsKspizero \nu_{\tau}$ 
and background modeled by MC.
The hatched histogram is the background from $\tau^{-}\to\piKsKs \nu_{\tau}$ and 
the shaded (yellow) histogram is the $q\bar{q}$ background.
See the text for details of the signal model for $\tauTO \piKsKspizero \nu_{\tau}$.}
\label{fig:h3200psubfull}
\end{figure*}

The invariant mass of the $\pi^{0}\Ks\Ks$ and $\Ks\pi^{-}$ subsystem for the 
$\tauTO \piKsKspizero \nu_{\tau}$ selected sample is shown 
in Fig.~\ref{fig:submass} (a) and (b), respectively.
The $M(\pi^{0}\Ks\Ks)$ distribution in Fig.~\ref{fig:submass} (a) 
shows a significant peak at 1280 MeV/$c^{2}$, 
which is probably due to the $\fone(1285)$ resonance.
In addition, a small bump-like structure is seen around 1420 MeV/$c^{2}$.
The $M(\Ks\pi^{-})$ distribution for the same 
$\tau^-\to \piKsKspizero \neutau$ sample, in 
Fig.~\ref{fig:submass} (b), shows a clear $K^{*}$ peak at 890 MeV/$c^2$.
These structures are also seen as clear bands in the two-dimensional plot,
$M(\Ks\pi^{-})$ versus  $M(\pi^{0}\Ks\Ks)$, as shown in Fig.~\ref{fig:submass} (c).
It should be noticed that no clear resonance-like structure is observed in the other sub-mass 
distributions as shown in Fig.~\ref{fig:h3200psubfull}.
In particular, there is no $\rho(770)$ signal in $M(\pi^{-}\pi^{0})$ and
no $K^{0*}$ signal in $M(\Ks\pi^{0})$. 
Altogether, this indicates the presence of two dominant components,
$\tau^{-} \to \pi^{-} f_{1}(1285) \nu_{\tau}$ and 
$\tau^{-} \to K^{*-}\Ks \pi^{0} \nu_{\tau}$,
in the final state of the decay $\tau^{-} \to \pi^{-} \Ks\Ks \pi^{0} \nu_{\tau}$.

In order to make a quantitative evaluation, 
we perform a simple amplitude analysis assuming incoherent contributions of
two intermediate processes $\tau^-\to \pi^-\fone(1285)\nu_{\tau}$ 
and $\tau^{-} \to K^{*-}\Ks \pi^{0} \nu_{\tau}$.
In addition, a possible contribution of $\fone(1420)$ production through 
$\tau^-\to \pi^-\fone(1420)\nu_{\tau}$ is also examined.

\subsection{Fitting formula}

We fit both the $M(\pi^{0}\Ks\Ks)$ and $M(\Ks\pi^{-})$ distributions 
in the decay 
$\tauTO \piKsKspizero \nu_{\tau}$ simultaneously, assuming that the dominant 
signal processes are those containing intermediate resonances 
$\fone(1285)$, $\fone(1420)$ and $\Kstar$, \textit{i.e.,} $\tau^-\to \pi^-\fone(1285)\nu_{\tau}$,  
$ \tau^-\to \pi^-\fone(1420)\nu_{\tau}$ and 
$\tau^{-} \to K^{*-}\Ks \pi^{0} \nu_{\tau}$.
Hereinafter, we refer to these decays as the $\fone$, $\foneprime$ and $\Kstar$ 
subprocesses, respectively.
 
We use an unbinned maximum-likelihood fit to extract the resonance parameters 
in the $M(\piKsKs)$ and $M(\Ks\pi^{-})$ distributions.
The likelihood function is given by
\begin{align}
\mathcal{L}=& \prod_{i}^{N} \left[ f_{\fone+\foneprime} \mathcal{P}_{\fone+\foneprime} (q_{1,i}, q_{2,i}; \vec{a})  
\right. \nonumber \\
  +&  \left.
   f_{\Kstar}\mathcal{P}_{\Kstar} (q_{1,i}, q_{2,i}; \vec{a})  \right. \nonumber \\ 
+& \left.
 (1 - f_{\fone+\foneprime} - f_{\Kstar})\mathcal{P}_{B} (q_{1,i}, q_{2,i}; \vec{a}) 
\right],
\end{align}
where $N$ is the total number of events in the sample, $f_{j}$ is the fraction of the $j$-th category, 
where the index $j$ stands for $\fone+\foneprime$, $\Kstar$ or the background ($B$) component.
$\mathcal{P}_{j}$ is the probability density function (PDF) for the $j$-th component.
The variables $q_{1,i}$ and $q_{2,i}$ are the invariant masses of the subsystems, 
\textit{i.e.,} $q_{1}=M(\piKsKs)$ and $q_{2}=M(\Ks\pi^{-})$, for the $i$-th event.
The vector $\vec{a}$ represents the resonance shape parameters.
We are aware of a possible interference between $\fone+\foneprime$ and $\Kstar$ amplitude; however, 
our statistics are too low for a quantitative study of this effect and so we ignore it in the fit.
We also assume that the PDF is given as the product of individual PDFs for each variables;
$\mathcal{P}_{j} = \mathcal{P}_{j}(q_{1}) \mathcal{P}_{j} (q_{2})$
for all components ($j=\fone + \foneprime, \Kstar,$ background).
As a result, we have six PDF's: 
$\mathcal{P}_{\fone +\foneprime} (q_{\alpha})$, $\mathcal{P}_{\Kstar} (q_{\alpha})$ and $\mathcal{P}_{B}(q_{\alpha})$ for $\alpha=1,2$.

The PDF $\mathcal{P}_{\fone +\foneprime} (q_{1})$ is the $M(\pi^{0}\Ks\Ks)$ distribution in the
$\tau^-\to \pi^-\fone(1285) (\fone(1420)) \nu_{\tau}$ decays and is given by
\begin{align}
\mathcal{P}_{\fone +\foneprime} (q_{1} ) \propto  
  | \beta~\textrm{BW}_{\fone(1285)}(q^{2}_{1})     \nonumber  \\ 
   + 
(1-\beta)~\textrm{BW}_{\fone(1420)}(q^{2}_{1}) |^{2},
\end{align}
where $\textrm{BW}_{X}(s)$ is the relativistic Breit-Wigner function and
$\beta$, a ratio of two resonances, is a real number.
$\textrm{BW}_{X}(s)$ is defined by
\begin{align}
\textrm{BW}_{X}(s) = \frac{\sqrt{s}M_{X}}{s-M^{2}_{X} + i\sqrt{s}\Gamma_{X}},
\label{eq:bw}
\end{align}
which describes the $\fone(1285)$ and $\fone(1420)$ resonance shape. 
$M_{X}$ and $\Gamma_{X}$ are the nominal mass and width for resonance $X$.

For the PDF $\mathcal{P}_{\Kstar} (q_{2})$, the Breit-Wigner function of 
Eq.~(\ref{eq:bw}) is used to describe the $K^{*-}$ resonance shape 
in the $M(\Ks\pi^{-})$ distribution:
$$\mathcal{P}_{\Kstar} (q_{2}) = |\textrm{BW}_{\Kstar}(q_{2}^{2}) |^{2}. $$ 
 
The PDF $\mathcal{P}_{\Kstar} (q_{1})$ is the $M( \pi^{0}\Ks\Ks)$ distribution 
for the $\tauTO K^{*-}\Ks\pi^{0} \nu_{\tau}$ decay.
In order to obtain this component, 
we generate $\tauTO K^{*-}\Ks\pi^{0} \nu_{\tau}( K^{*-}\to \Ks \pi^{-})$ events 
using {\tt PYTHIA} 6.4~\cite{Sjostrand:2006za}, assuming phase space 
for the $K^{*-}\Ks \pi^{0}$ system 
(see the $K^{*-} \Ks \pi^{0}$ contribution in Fig.~\ref{fig:submass}(a)).
Note that this distribution is insensitive to the detailed values 
of the $K^{*}$ resonance parameters.
The two-dimensional plot, $M(\pi^{0}\Ks\Ks)$ versus $M(\Ks\pi^{-})$, 
for the $\tau^{-} \to \Kstar\Ks\pi^{0}\nu_{\tau}$
MC events is shown in Fig.~\ref{fig:submass}(d).

The PDF $\mathcal{P}_{\fone+\foneprime} (q_{2})$ is the  $M(\Ks\pi^{-})$ 
distribution for the $ \tau^-\to \pi^-\fone(1285) ( \fone(1420)) \nu_{\tau}$ 
decays.
In order to obtain this component, 
we generate $\tau^-\to \pi^-\fone(1285) \nu_{\tau}~ (\fone(1285)\to \Ks\Ks\pi^{0})$ 
events with the {\tt PYTHIA} 6.4 code~\cite{Sjostrand:2006za} and obtain 
the shape of the $M(\Ks\pi^{-})$ distribution (see the 
$\fone(1285)\pi^{-} + \fone(1420)\pi^{-} $ contribution in 
Fig.~\ref{fig:submass}(b)).
The same two-dimensional plot for the $\tau^{-} \to \pi^{-}\fone(1285)\nu_{\tau} + \pi^{-}\fone(1420) \nu_{\tau}$
is shown in Fig.~\ref{fig:submass}(e).

The dominant background for the $\tau^-\to \piKsKspizero \nu_{\tau}$ sample 
is due to the $\tau^-\to \piKsKs \nu_{\tau}$ decay with a fake $\pi^{0}$. 
In addition, there is a small contribution from the $q\bar{q}$ continuum.
In order to model the background component,
we tune the mass distribution of the $\tau^-\to \piKsKs \nu_{\tau}$ 
MC events to agree with the data.
The background PDF $\mathcal{P}_{B}(q_{i})$,  prepared from 
the MC prediction, is shown by the shaded histograms of 
$M(\piKsKs)$ and $M(\Ks\pi^{-})$ in Fig.~\ref{fig:submass}(a) and (b), respectively.

\subsection{Fit results}

The fit results with
$\fone(1285)\pi^- \nu_{\tau}$, $\fone(1420)\pi^- \nu_{\tau}$, 
$K^{*-}\Ks\pi^{0} \nu_{\tau}$
and background contributions reproduce the data quite 
well as shown by the solid line in Fig.~\ref{fig:submass} (a) and (b).

The significance of the $\fone(1420)$ component is obtained from 
the negative log-likelihood difference with and without 
the $\fone(1420)$ signal, 
$S=-2 \rm{ln}(\mathcal{L}_{0}/\mathcal{L}_{\rm max})$, 
where $\mathcal{L}_{\rm max}$ and $\mathcal{L}_{0}$ is the likelihood with and without 
the $\fone(1420)$ resonance, respectively.
We obtain $S=30$ with a change of the number of degrees of freedom by 3.
From these results, we conclude that the significance of the $\fone(1420)$ 
is $4.8\sigma$.
In the same way, the significances of $\fone(1285)$ and $K^{*-}$ are 
$12\sigma$ and $7.8\sigma$, respectively.

As a result of the fit, the masses and widths for the $\fone(1285)$, $\fone(1420)$ 
and $K^{*-}$ are determined to be
\begin{align}
m_{\fone(1285)} = 1274 \pm 3 ~\textrm{MeV}/c^{2},    \nonumber \\ 
\Gamma_{\fone(1285)} = 20 \pm 4 ~\textrm{MeV}/c^{2},  \nonumber \\
m_{\fone(1420)} = 1425 \pm 2 ~\textrm{MeV}/c^{2},     \nonumber \\
\Gamma_{\fone(1420)} = 42 \pm 19 ~\textrm{MeV}/c^{2}.  \nonumber \\
m_{K^{*-}} =  890 \pm 3 ~\textrm{MeV}/c^{2},    \nonumber \\ 
\Gamma_{K^{*-}} = 48 \pm 2 ~\textrm{MeV}/c^{2}.  \nonumber 
\end{align}
These results are consistent with the world averages~\cite{Beringer:1900zz}.

The fractions of the three hadronic currents in 
$\tauTO \piKsKspizero \nu_{\tau}$ 
are determined to be $(34 \pm 5)\%$, $(12 \pm 3)\%$ and $(54 \pm 6)\%$
for the $\fone(1285)\pi^{-}\nu_{\tau}$, $\fone(1420)\pi^{-}\nu_{\tau}$ 
and $K^{*-}\Ks\pi^{0}\nu_{\tau}$ modes, respectively.

Using the fraction of each component, the products of the branching 
fractions for the subprocesses are determined to be
\begin{align}
 \Br(\tauTO \fone(1285) \pi^{-} \nu_{\tau}) \cdot \Br(\fone(1285) \to \Ks\Ks\pi^{0}) \nonumber \\  
   = (0.68 \pm 0.13 \pm 0.07) \times 10^{-5},  \nonumber  \\
 \Br(\tauTO \fone(1420) \pi^{-} \nu_{\tau}) \cdot \Br(\fone(1420) \to \Ks\Ks\pi^{0}) \nonumber \\  
   = (0.24 \pm 0.05 \pm 0.06) \times 10^{-5}, \nonumber  \\
 \Br(\tauTO K^{*-} \Ks \pi^{0} \nu_{\tau}) \cdot \Br(\Kstar\to \Ks \pi^-) \nonumber \\ 
   = (1.08 \pm 0.14 \pm 0.15) \times 10^{-5}.  \nonumber
\end{align}
The first uncertainty is statistical and the second is systematic.
The systematic uncertainties are estimated by using different fit methods, 
such as a 1-D fit and a simultaneous fit of two sub-mass distributions.
Both statistical and systematical uncertainties of 
$\Br(\tauTO \piKsKspizero \nu_{\tau})$ are taken into account as well.

In addition, we examined other subsystems by generating MC events 
with the ratios of three processes obtained by the above-mentioned fit 
shown in Fig.~\ref{fig:h3200psubfull}.
The blank histograms (red), the sum of the $\fone(1285)\pi^{-}\nu_{\tau}$, 
$\fone(1420)\pi^{-}\nu_{\tau}$, and $K^{*-}\Ks\pi^{0}\nu_{\tau}$ processes and the other backgrounds, 
show the expected distributions of the invariant masses of 
the other subsystems in the $\tauTO \piKsKspizero \nu_{\tau}$ sample.
We use the shape of these three processes obtained by {\tt PYTHIA} 6.4~\cite{Sjostrand:2006za} 
and the fit results for the relative ratio of these components.
A small contribution due to the interference between the $\fone(1285)$ and $\fone(1420)$
resonances is ignored.
The invariant mass distributions of all subsystems are explained by 
this model quite well in our data.

\section{Conclusions}
\label{sec:conclusion}
Using $616 \times 10^{6}$ $\taupm$ events collected with the Belle detector,
we measure the inclusive $\Ks$ and six exclusive branching fractions and the covariance matrix
for hadronic decays of the $\tau$ lepton containing $\Ks$: 
  $\piKs\nu_{\tau}$, $\KKs\nu_{\tau}$, $\piKspizero\nu_{\tau}$, 
  $\KKspizero\nu_{\tau}$, $\piKsKs\nu_{\tau}$ and $\piKsKspizero\nu_{\tau}$.
Our results are summarized in Table~\ref{tab:results}.
The result for $\taum \to \piKs \nu_{\tau}$ supersedes our previous measurement~\cite{Epifanov:2007rf}.
The accuracy for $\taum \to \KKs \nu_{\tau}$, $\taum \to \piKspizero \nu_{\tau}$ and 
$\taum \to \KKspizero \nu_{\tau}$ is improved over that of 
previous experiments by one order of magnitude.

The combined fit of the invariant masses of the $\pi^{0}\Ks\Ks$ 
and $\pi^{-}\Ks$ system in the $\tauTO \piKsKspizero \nu_{\tau}$ events 
indicates the presence of the $\Kstar \Ks \pi^{0}\neutau$, $\fone(1285)\pi^{-}\neutau$ and 
$\fone(1420)\pi^{-}\neutau$ components with  significances of 7.8$\sigma$, 
12$\sigma$ and 4.8$\sigma$, respectively.
Using the branching fractions of the intermediate resonances to the 
corresponding final states from \cite{Beringer:1900zz},
the branching fractions for the $\piKsKspizero \nu_{\tau}$ 
final state via hadronic currents are determined to be
$\Br(\tauTO \fone(1285) \pi^{-} \nu_{\tau}) = (0.68 \pm 0.13 \pm 0.07) \times 10^{-5}$,
$\Br(\tauTO \fone(1420) \pi^{-} \nu_{\tau}) = (0.24 \pm 0.05 \pm 0.06) \times 10^{-5}$,
and $\Br(\tauTO K^{*-} \Ks \pi^{0} \nu_{\tau})  = (1.08 \pm 0.14 \pm 0.15) \times 10^{-5}$.

\section{Acknowledgment}
%

We thank the KEKB group for the excellent operation of the
accelerator; the KEK cryogenics group for the efficient
operation of the solenoid; and the KEK computer group,
the National Institute of Informatics, and the 
PNNL/EMSL computing group for valuable computing
and SINET4 network support.  We acknowledge support from
the Ministry of Education, Culture, Sports, Science, and
Technology (MEXT) of Japan, the Japan Society for the 
Promotion of Science (JSPS), and the Tau-Lepton Physics 
Research Center of Nagoya University; 
the Australian Research Council and the Australian 
Department of Industry, Innovation, Science and Research;
Austrian Science Fund under Grant No. P 22742-N16;
the National Natural Science Foundation of China under contract 
No.~10575109, 10775142, 10825524, 10875115, 10935008 and 11175187; 
the Ministry of Education, Youth and Sports of the Czech 
Republic under contract No.~MSM0021620859;
the Carl Zeiss Foundation, the Deutsche Forschungsgemeinschaft
and the VolkswagenStiftung;
the Department of Science and Technology of India; 
the Istituto Nazionale di Fisica Nucleare of Italy; 
The WCU program of the Ministry Education Science and
Technology, National Research Foundation of Korea Grant No.
2011-0029457, 2012-0008143, 2012R1A1A2008330, 2013R1A1A3007772,
BRL program under NRF Grant No. KRF-2011-0020333, BK21 Plus program,
and GSDC of the Korea Institute of Science and Technology Information;
the Polish Ministry of Science and Higher Education and 
the National Science Center;
the Ministry of Education and Science of the Russian Federation, 
the Russian Federal Agency for Atomic Energy and the RFBR grant 12-02-01032-a;
the Slovenian Research Agency;
the Basque Foundation for Science (IKERBASQUE) and the UPV/EHU under 
program UFI 11/55;
the Swiss National Science Foundation; the National Science Council
and the Ministry of Education of Taiwan; and the U.S.\
Department of Energy and the National Science Foundation.
This work is supported by a Grant-in-Aid from MEXT for 
Science Research in a Priority Area (``New Development of 
Flavor Physics''), and from JSPS for Creative Scientific 
Research (``Evolution of Tau-lepton Physics'').


\newpage
\section*{Appendix}
\label{lab:appendix}
In this appendix, we provide the description of the logarithmic Gaussian 
that is used to model the $\gamma\gamma$ invariant mass distribution, 
which is inadequately described with a pure Gaussian distribution.
This function is useful for modeling the distribution that has an asymmetrical tail.
The normalized logarithmic Gaussian $f(x)$ is given by
\begin{equation}
f(x)=\frac{\eta}{\sqrt{2\pi}\sigma \sigma_{0}} \exp \left(
 -  \frac{\ln^{2}\left( 1- \eta(x - x_{p})/\sigma \right)  }{2\sigma_{0}^{2}}
  -\frac{\sigma^{2}_{0}}{2}   \right) , 
\end{equation}
where $x_{p}$, $\sigma$ and $\eta$ are  free parameters in this function. 
The parameter $x_{p}$ represents  the peak position, $\sigma$ characterizes the mean standard deviation
 of the distribution and
$\eta$ represents the asymmetry of the distribution. 
As $\eta$ approaches zero, this distribution collapses
to a Gaussian.
The variable $\sigma_{0}$ is determined by $\eta$ as
\begin{eqnarray}
\sigma_{0} =\frac{2}{\xi}\sinh^{-1}\left( \frac{\eta \xi}{2} \right)
 \end{eqnarray}    
 with $\xi = 2 \sqrt{\ln{4}}\sim 2.35$. 
The left and right standard deviation ($\sigma_{\pm}$) and 
the $x$-values ($X_{\pm}$) for which the distribution decreased 
by a factor of $P$ from the value at the maximum of the distribution are given by 
 \begin{eqnarray}
 \sigma_{\pm} &=& \pm \frac{\sigma}{\eta} \left(  1 - e^{\mp \frac{\sigma_{0}\xi}{2}} \right)  \\
 X_{\pm} &=& x_{p} + \frac{\sigma}{\eta} \left(  1 - e^{\mp \sigma_{0}\sqrt{2\ln P}} \right).
\end{eqnarray}


\bibliography{inc.bib}

\begin{thebibliography}{39}%
\makeatletter
\providecommand \@ifxundefined [1]{%
 \@ifx{#1\undefined}
}%
\providecommand \@ifnum [1]{%
 \ifnum #1\expandafter \@firstoftwo
 \else \expandafter \@secondoftwo
 \fi
}%
\providecommand \@ifx [1]{%
 \ifx #1\expandafter \@firstoftwo
 \else \expandafter \@secondoftwo
 \fi
}%
\providecommand \natexlab [1]{#1}%
\providecommand \enquote  [1]{``#1''}%
\providecommand \bibnamefont  [1]{#1}%
\providecommand \bibfnamefont [1]{#1}%
\providecommand \citenamefont [1]{#1}%
\providecommand \href@noop [0]{\@secondoftwo}%
\providecommand \href [0]{\begingroup \@sanitize@url \@href}%
\providecommand \@href[1]{\@@startlink{#1}\@@href}%
\providecommand \@@href[1]{\endgroup#1\@@endlink}%
\providecommand \@sanitize@url [0]{\catcode `\\12\catcode `\$12\catcode
  `\&12\catcode `\#12\catcode `\^12\catcode `\_12\catcode `\%12\relax}%
\providecommand \@@startlink[1]{}%
\providecommand \@@endlink[0]{}%
\providecommand \url  [0]{\begingroup\@sanitize@url \@url }%
\providecommand \@url [1]{\endgroup\@href {#1}{\urlprefix }}%
\providecommand \urlprefix  [0]{URL }%
\providecommand \Eprint [0]{\href }%
\providecommand \doibase [0]{http://dx.doi.org/}%
\providecommand \selectlanguage [0]{\@gobble}%
\providecommand \bibinfo  [0]{\@secondoftwo}%
\providecommand \bibfield  [0]{\@secondoftwo}%
\providecommand \translation [1]{[#1]}%
\providecommand \BibitemOpen [0]{}%
\providecommand \bibitemStop [0]{}%
\providecommand \bibitemNoStop [0]{.\EOS\space}%
\providecommand \EOS [0]{\spacefactor3000\relax}%
\providecommand \BibitemShut  [1]{\csname bibitem#1\endcsname}%
\let\auto@bib@innerbib\@empty
\bibitem [{\citenamefont {Narison}\ and\ \citenamefont
  {Pich}(1988)}]{Narison:1988ni}%
  \BibitemOpen
  \bibfield  {author} {\bibinfo {author} {\bibfnamefont {S.}~\bibnamefont
  {Narison}}\ and\ \bibinfo {author} {\bibfnamefont {A.}~\bibnamefont {Pich}},\
  }\href {\doibase 10.1016/0370-2693(88)90830-1} {\bibfield  {journal}
  {\bibinfo  {journal} {Phys. Lett. B}\ }\textbf {\bibinfo {volume} {211}},\
  \bibinfo {pages} {183} (\bibinfo {year} {1988})}\BibitemShut {NoStop}%
\bibitem [{\citenamefont {Schael}\ \emph {et~al.}(2005)\citenamefont {Schael}
  \emph {et~al.}}]{Schael:2005am}%
  \BibitemOpen
  \bibfield  {author} {\bibinfo {author} {\bibfnamefont {S.}~\bibnamefont
  {Schael}} \emph {et~al.} (\bibinfo {collaboration} {ALEPH Collaboration}),\
  }\href {\doibase 10.1016/j.physrep.2005.06.007} {\bibfield  {journal}
  {\bibinfo  {journal} {Phys. Rept.}\ }\textbf {\bibinfo {volume} {421}},\
  \bibinfo {pages} {191} (\bibinfo {year} {2005})},\ \Eprint
  {http://arxiv.org/abs/hep-ex/0506072} {arXiv:hep-ex/0506072} \BibitemShut
  {NoStop}%
\bibitem [{\citenamefont {Ackerstaff}\ \emph {et~al.}(1999)\citenamefont
  {Ackerstaff} \emph {et~al.}}]{Ackerstaff:1998yj}%
  \BibitemOpen
  \bibfield  {author} {\bibinfo {author} {\bibfnamefont {K.}~\bibnamefont
  {Ackerstaff}} \emph {et~al.} (\bibinfo {collaboration} {OPAL
  Collaboration}),\ }\href {\doibase 10.1007/s100529901061} {\bibfield
  {journal} {\bibinfo  {journal} {Eur. Phys. J. C}\ }\textbf {\bibinfo {volume}
  {7}},\ \bibinfo {pages} {571} (\bibinfo {year} {1999})},\ \Eprint
  {http://arxiv.org/abs/hep-ex/9808019} {arXiv:hep-ex/9808019} \BibitemShut
  {NoStop}%
\bibitem [{\citenamefont {Gamiz}\ \emph {et~al.}(2005)\citenamefont {Gamiz},
  \citenamefont {Jamin}, \citenamefont {Pich}, \citenamefont {Prades},\ and\
  \citenamefont {Schwab}}]{Gamiz:2004ar}%
  \BibitemOpen
  \bibfield  {author} {\bibinfo {author} {\bibfnamefont {E.}~\bibnamefont
  {Gamiz}}, \bibinfo {author} {\bibfnamefont {M.}~\bibnamefont {Jamin}},
  \bibinfo {author} {\bibfnamefont {A.}~\bibnamefont {Pich}}, \bibinfo {author}
  {\bibfnamefont {J.}~\bibnamefont {Prades}}, \ and\ \bibinfo {author}
  {\bibfnamefont {F.}~\bibnamefont {Schwab}},\ }\href {\doibase
  10.1103/PhysRevLett.94.011803} {\bibfield  {journal} {\bibinfo  {journal}
  {Phys. Rev. Lett.}\ }\textbf {\bibinfo {volume} {94}},\ \bibinfo {pages}
  {011803} (\bibinfo {year} {2005})},\ \Eprint
  {http://arxiv.org/abs/hep-ph/0408044} {arXiv:hep-ph/0408044} \BibitemShut
  {NoStop}%
\bibitem [{\citenamefont {Baikov}\ \emph {et~al.}(2005)\citenamefont {Baikov},
  \citenamefont {Chetyrkin},\ and\ \citenamefont {K\"uhn}}]{Baikov:2004tk}%
  \BibitemOpen
  \bibfield  {author} {\bibinfo {author} {\bibfnamefont {P.}~\bibnamefont
  {Baikov}}, \bibinfo {author} {\bibfnamefont {K.}~\bibnamefont {Chetyrkin}}, \
  and\ \bibinfo {author} {\bibfnamefont {J.~H.}\ \bibnamefont {K\"uhn}},\
  }\href {\doibase 10.1103/PhysRevLett.95.012003} {\bibfield  {journal}
  {\bibinfo  {journal} {Phys. Rev. Lett.}\ }\textbf {\bibinfo {volume} {95}},\
  \bibinfo {pages} {012003} (\bibinfo {year} {2005})},\ \Eprint
  {http://arxiv.org/abs/hep-ph/0412350} {arXiv:hep-ph/0412350} \BibitemShut
  {NoStop}%
\bibitem [{\citenamefont {Kambor}\ and\ \citenamefont
  {Maltman}(2000)}]{Kambor:2000dj}%
  \BibitemOpen
  \bibfield  {author} {\bibinfo {author} {\bibfnamefont {J.}~\bibnamefont
  {Kambor}}\ and\ \bibinfo {author} {\bibfnamefont {K.}~\bibnamefont
  {Maltman}},\ }\href {\doibase 10.1103/PhysRevD.62.093023} {\bibfield
  {journal} {\bibinfo  {journal} {Phys. Rev. D}\ }\textbf {\bibinfo {volume}
  {62}},\ \bibinfo {pages} {093023} (\bibinfo {year} {2000})},\ \Eprint
  {http://arxiv.org/abs/hep-ph/0005156} {arXiv:hep-ph/0005156} \BibitemShut
  {NoStop}%
\bibitem [{\citenamefont {Barate}\ \emph
  {et~al.}(1999{\natexlab{a}})\citenamefont {Barate} \emph
  {et~al.}}]{Barate:1999hi}%
  \BibitemOpen
  \bibfield  {author} {\bibinfo {author} {\bibfnamefont {R.}~\bibnamefont
  {Barate}} \emph {et~al.} (\bibinfo {collaboration} {ALEPH Collaboration}),\
  }\href {\doibase 10.1007/s100529900146} {\bibfield  {journal} {\bibinfo
  {journal} {Eur. Phys. J. C}\ }\textbf {\bibinfo {volume} {10}},\ \bibinfo
  {pages} {1} (\bibinfo {year} {1999}{\natexlab{a}})},\ \Eprint
  {http://arxiv.org/abs/hep-ex/9903014} {arXiv:hep-ex/9903014} \BibitemShut
  {NoStop}%
\bibitem [{\citenamefont {Barate}\ \emph
  {et~al.}(1999{\natexlab{b}})\citenamefont {Barate} \emph
  {et~al.}}]{Barate:1999hj}%
  \BibitemOpen
  \bibfield  {author} {\bibinfo {author} {\bibfnamefont {R.}~\bibnamefont
  {Barate}} \emph {et~al.} (\bibinfo {collaboration} {ALEPH Collaboration}),\
  }\href {\doibase 10.1007/s100520050659} {\bibfield  {journal} {\bibinfo
  {journal} {Eur. Phys. J. C}\ }\textbf {\bibinfo {volume} {11}},\ \bibinfo
  {pages} {599} (\bibinfo {year} {1999}{\natexlab{b}})},\ \Eprint
  {http://arxiv.org/abs/hep-ex/9903015} {arXiv:hep-ex/9903015} \BibitemShut
  {NoStop}%
\bibitem [{\citenamefont {Abbiendi}\ \emph {et~al.}(2000)\citenamefont
  {Abbiendi} \emph {et~al.}}]{Abbiendi:1999pm}%
  \BibitemOpen
  \bibfield  {author} {\bibinfo {author} {\bibfnamefont {G.}~\bibnamefont
  {Abbiendi}} \emph {et~al.} (\bibinfo {collaboration} {OPAL Collaboration}),\
  }\href {\doibase 10.1007/s100520000317} {\bibfield  {journal} {\bibinfo
  {journal} {Eur. Phys. J. C}\ }\textbf {\bibinfo {volume} {13}},\ \bibinfo
  {pages} {213} (\bibinfo {year} {2000})},\ \Eprint
  {http://arxiv.org/abs/hep-ex/9911029} {arXiv:hep-ex/9911029} \BibitemShut
  {NoStop}%
\bibitem [{\citenamefont {Coan}\ \emph {et~al.}(1996)\citenamefont {Coan} \emph
  {et~al.}}]{PhysRevD.53.6037}%
  \BibitemOpen
  \bibfield  {author} {\bibinfo {author} {\bibfnamefont {T.~E.}\ \bibnamefont
  {Coan}} \emph {et~al.} (\bibinfo {collaboration} {CLEO Collaboration}),\
  }\href {\doibase 10.1103/PhysRevD.53.6037} {\bibfield  {journal} {\bibinfo
  {journal} {Phys. Rev. D}\ }\textbf {\bibinfo {volume} {53}},\ \bibinfo
  {pages} {6037} (\bibinfo {year} {1996})}\BibitemShut {NoStop}%
\bibitem [{\citenamefont {Epifanov}\ \emph {et~al.}(2007)\citenamefont
  {Epifanov} \emph {et~al.}}]{Epifanov:2007rf}%
  \BibitemOpen
  \bibfield  {author} {\bibinfo {author} {\bibfnamefont {D.}~\bibnamefont
  {Epifanov}} \emph {et~al.} (\bibinfo {collaboration} {Belle Collaboration}),\
  }\href {\doibase 10.1016/j.physletb.2007.08.045} {\bibfield  {journal}
  {\bibinfo  {journal} {Phys. Lett. B}\ }\textbf {\bibinfo {volume} {654}},\
  \bibinfo {pages} {65} (\bibinfo {year} {2007})},\ \Eprint
  {http://arxiv.org/abs/0706.2231} {arXiv:0706.2231} \BibitemShut {NoStop}%
\bibitem [{\citenamefont {Aubert}\ \emph {et~al.}(2007)\citenamefont {Aubert}
  \emph {et~al.}}]{Aubert:2007jh}%
  \BibitemOpen
  \bibfield  {author} {\bibinfo {author} {\bibfnamefont {B.}~\bibnamefont
  {Aubert}} \emph {et~al.} (\bibinfo {collaboration} {BaBar Collaboration}),\
  }\href {\doibase 10.1103/PhysRevD.76.051104} {\bibfield  {journal} {\bibinfo
  {journal} {Phys. Rev. D}\ }\textbf {\bibinfo {volume} {76}},\ \bibinfo
  {pages} {051104} (\bibinfo {year} {2007})},\ \Eprint
  {http://arxiv.org/abs/0707.2922} {arXiv:0707.2922} \BibitemShut {NoStop}%
\bibitem [{\citenamefont {Lee}\ \emph {et~al.}(2010)\citenamefont {Lee} \emph
  {et~al.}}]{Lee:2010tc}%
  \BibitemOpen
  \bibfield  {author} {\bibinfo {author} {\bibfnamefont {M.~J.}\ \bibnamefont
  {Lee}} \emph {et~al.} (\bibinfo {collaboration} {Belle Collaboration}),\
  }\href {\doibase 10.1103/PhysRevD.81.113007} {\bibfield  {journal} {\bibinfo
  {journal} {Phys. Rev. D}\ }\textbf {\bibinfo {volume} {81}},\ \bibinfo
  {pages} {113007} (\bibinfo {year} {2010})},\ \Eprint
  {http://arxiv.org/abs/1001.0083} {arXiv:1001.0083} \BibitemShut {NoStop}%
\bibitem [{\citenamefont {Aubert}\ \emph {et~al.}(2010)\citenamefont {Aubert}
  \emph {et~al.}}]{Aubert:2009qj}%
  \BibitemOpen
  \bibfield  {author} {\bibinfo {author} {\bibfnamefont {B.}~\bibnamefont
  {Aubert}} \emph {et~al.} (\bibinfo {collaboration} {BaBar Collaboration}),\
  }\href {\doibase 10.1103/PhysRevLett.105.051602} {\bibfield  {journal}
  {\bibinfo  {journal} {Phys. Rev. Lett.}\ }\textbf {\bibinfo {volume} {105}},\
  \bibinfo {pages} {051602} (\bibinfo {year} {2010})},\ \Eprint
  {http://arxiv.org/abs/0912.0242} {arXiv:0912.0242} \BibitemShut {NoStop}%
\bibitem [{\citenamefont {Aubert}\ \emph {et~al.}(2008)\citenamefont {Aubert}
  \emph {et~al.}}]{Aubert:2007mh}%
  \BibitemOpen
  \bibfield  {author} {\bibinfo {author} {\bibfnamefont {B.}~\bibnamefont
  {Aubert}} \emph {et~al.} (\bibinfo {collaboration} {BaBar Collaboration}),\
  }\href {\doibase 10.1103/PhysRevLett.100.011801} {\bibfield  {journal}
  {\bibinfo  {journal} {Phys. Rev. Lett.}\ }\textbf {\bibinfo {volume} {100}},\
  \bibinfo {pages} {011801} (\bibinfo {year} {2008})},\ \Eprint
  {http://arxiv.org/abs/0707.2981} {arXiv:0707.2981} \BibitemShut {NoStop}%
\bibitem [{\citenamefont {Inami}\ \emph {et~al.}(2009)\citenamefont {Inami}
  \emph {et~al.}}]{Inami:2008ar}%
  \BibitemOpen
  \bibfield  {author} {\bibinfo {author} {\bibfnamefont {K.}~\bibnamefont
  {Inami}} \emph {et~al.} (\bibinfo {collaboration} {Belle Collaboration}),\
  }\href {\doibase 10.1016/j.physletb.2009.01.047} {\bibfield  {journal}
  {\bibinfo  {journal} {Phys. Lett. B}\ }\textbf {\bibinfo {volume} {672}},\
  \bibinfo {pages} {209} (\bibinfo {year} {2009})},\ \Eprint
  {http://arxiv.org/abs/0811.0088} {arXiv:0811.0088} \BibitemShut {NoStop}%
\bibitem [{\citenamefont {del Amo~Sanchez}\ \emph {et~al.}(2011)\citenamefont
  {del Amo~Sanchez} \emph {et~al.}}]{delAmoSanchez:2010pc}%
  \BibitemOpen
  \bibfield  {author} {\bibinfo {author} {\bibfnamefont {P.}~\bibnamefont {del
  Amo~Sanchez}} \emph {et~al.} (\bibinfo {collaboration} {BaBar
  Collaboration}),\ }\href {\doibase 10.1103/PhysRevD.83.032002} {\bibfield
  {journal} {\bibinfo  {journal} {Phys. Rev. D}\ }\textbf {\bibinfo {volume}
  {83}},\ \bibinfo {pages} {032002} (\bibinfo {year} {2011})},\ \Eprint
  {http://arxiv.org/abs/1011.3917} {arXiv:1011.3917} \BibitemShut {NoStop}%
\bibitem [{\citenamefont {Lees}\ \emph {et~al.}(2012)\citenamefont {Lees} \emph
  {et~al.}}]{Lees:2012de}%
  \BibitemOpen
  \bibfield  {author} {\bibinfo {author} {\bibfnamefont {J.~P.}\ \bibnamefont
  {Lees}} \emph {et~al.} (\bibinfo {collaboration} {BaBar Collaboration}),\
  }\href {\doibase 10.1103/PhysRevD.86.092013} {\bibfield  {journal} {\bibinfo
  {journal} {Phys. Rev. D}\ }\textbf {\bibinfo {volume} {86}},\ \bibinfo
  {pages} {092013} (\bibinfo {year} {2012})},\ \Eprint
  {http://arxiv.org/abs/1208.0376} {arXiv:1208.0376} \BibitemShut {NoStop}%
\bibitem [{\citenamefont {Kurokawa}\ and\ \citenamefont
  {Kikutani}(2003)}]{Kurokawa:2003io}%
  \BibitemOpen
  \bibfield  {author} {\bibinfo {author} {\bibfnamefont {S.}~\bibnamefont
  {Kurokawa}}\ and\ \bibinfo {author} {\bibfnamefont {E.}~\bibnamefont
  {Kikutani}},\ }\href@noop {} {\bibfield  {journal} {\bibinfo  {journal}
  {Nucl. Instrum. and Meth. A}\ }\textbf {\bibinfo {volume} {499}},\ \bibinfo
  {pages} {1} (\bibinfo {year} {2003})},\ \bibinfo {note} {and other papers
  included in this volume}\BibitemShut {NoStop}%
\bibitem [{\citenamefont {Abe}\ \emph {et~al.}(2013)\citenamefont {Abe} \emph
  {et~al.}}]{Abe2013aa}%
  \BibitemOpen
  \bibfield  {author} {\bibinfo {author} {\bibfnamefont {T.}~\bibnamefont
  {Abe}} \emph {et~al.},\ }\href {\doibase 10.1093/ptep/pts102} {\bibfield
  {journal} {\bibinfo  {journal} {Prog. Theor. Exp. Phys.}\ }\textbf {\bibinfo
  {volume} {3}},\ \bibinfo {pages} {03A001} (\bibinfo {year} {2013})},\
  \bibinfo {note} {and following articles up to 03A011}\BibitemShut {NoStop}%
\bibitem [{\citenamefont {Natkaniec}\ \emph {et~al.}(2006)\citenamefont
  {Natkaniec}, \citenamefont {Aihara}, \citenamefont {Asano}, \citenamefont
  {Aso}, \citenamefont {Bakich} \emph {et~al.}}]{Natkaniec:2006rv}%
  \BibitemOpen
  \bibfield  {author} {\bibinfo {author} {\bibfnamefont {Z.}~\bibnamefont
  {Natkaniec}}, \bibinfo {author} {\bibfnamefont {H.}~\bibnamefont {Aihara}},
  \bibinfo {author} {\bibfnamefont {Y.}~\bibnamefont {Asano}}, \bibinfo
  {author} {\bibfnamefont {T.}~\bibnamefont {Aso}}, \bibinfo {author}
  {\bibfnamefont {A.}~\bibnamefont {Bakich}},  \emph {et~al.},\ }\href
  {\doibase 10.1016/j.nima.2005.11.228} {\bibfield  {journal} {\bibinfo
  {journal} {Nucl. Instrum. and. Meth. A}\ }\textbf {\bibinfo {volume} {560}},\
  \bibinfo {pages} {1} (\bibinfo {year} {2006})}\BibitemShut {NoStop}%
\bibitem [{\citenamefont {Abashian}\ \emph
  {et~al.}(2002{\natexlab{a}})\citenamefont {Abashian} \emph
  {et~al.}}]{Abashian2002117}%
  \BibitemOpen
  \bibfield  {author} {\bibinfo {author} {\bibfnamefont {A.}~\bibnamefont
  {Abashian}} \emph {et~al.} (\bibinfo {collaboration} {Belle Collaboration}),\
  }\href {\doibase 10.1016/S0168-9002(01)02013-7} {\bibfield  {journal}
  {\bibinfo  {journal} {Nucl. Instrum. and. Meth. A}\ }\textbf {\bibinfo
  {volume} {479}},\ \bibinfo {pages} {117 } (\bibinfo {year}
  {2002}{\natexlab{a}})}\BibitemShut {NoStop}%
\bibitem [{\citenamefont {Brodzicka}\ \emph {et~al.}(2012)\citenamefont
  {Brodzicka} \emph {et~al.}}]{Brodzicka2012aa}%
  \BibitemOpen
  \bibfield  {author} {\bibinfo {author} {\bibfnamefont {J.}~\bibnamefont
  {Brodzicka}} \emph {et~al.},\ }\href {\doibase 10.1093/ptep/pts072}
  {\bibfield  {journal} {\bibinfo  {journal} {Prog. Theor. Exp. Phys.}\
  }\textbf {\bibinfo {volume} {1}},\ \bibinfo {pages} {04D001} (\bibinfo {year}
  {2012})},\ \bibinfo {note} {see detector section}\BibitemShut {NoStop}%
\bibitem [{\citenamefont {Jadach}\ \emph {et~al.}(2000)\citenamefont {Jadach},
  \citenamefont {Ward},\ and\ \citenamefont {W\c{a}s}}]{Jadach:1999vf}%
  \BibitemOpen
  \bibfield  {author} {\bibinfo {author} {\bibfnamefont {S.}~\bibnamefont
  {Jadach}}, \bibinfo {author} {\bibfnamefont {B.}~\bibnamefont {Ward}}, \ and\
  \bibinfo {author} {\bibfnamefont {Z.}~\bibnamefont {W\c{a}s}},\ }\href
  {\doibase 10.1016/S0010-4655(00)00048-5} {\bibfield  {journal} {\bibinfo
  {journal} {Comput. Phys. Commun.}\ }\textbf {\bibinfo {volume} {130}},\
  \bibinfo {pages} {260} (\bibinfo {year} {2000})},\ \Eprint
  {http://arxiv.org/abs/hep-ph/9912214} {arXiv:hep-ph/9912214} \BibitemShut
  {NoStop}%
\bibitem [{\citenamefont {Jadach}\ \emph {et~al.}(1993)\citenamefont {Jadach},
  \citenamefont {Was}, \citenamefont {Decker},\ and\ \citenamefont
  {Kuhn}}]{Jadach:1993hs}%
  \BibitemOpen
  \bibfield  {author} {\bibinfo {author} {\bibfnamefont {S.}~\bibnamefont
  {Jadach}}, \bibinfo {author} {\bibfnamefont {Z.}~\bibnamefont {Was}},
  \bibinfo {author} {\bibfnamefont {R.}~\bibnamefont {Decker}}, \ and\ \bibinfo
  {author} {\bibfnamefont {J.~H.}\ \bibnamefont {Kuhn}},\ }\href {\doibase
  10.1016/0010-4655(93)90061-G} {\bibfield  {journal} {\bibinfo  {journal}
  {Comput. Phys. Commun.}\ }\textbf {\bibinfo {volume} {76}},\ \bibinfo {pages}
  {361} (\bibinfo {year} {1993})}\BibitemShut {NoStop}%
\bibitem [{\citenamefont {Golonka}\ \emph {et~al.}(2006)\citenamefont
  {Golonka}, \citenamefont {Kersevan}, \citenamefont {Pierzchala},
  \citenamefont {Richter-W\c{a}s}, \citenamefont {W\c{a}s} \emph
  {et~al.}}]{Golonka:2003xt}%
  \BibitemOpen
  \bibfield  {author} {\bibinfo {author} {\bibfnamefont {P.}~\bibnamefont
  {Golonka}}, \bibinfo {author} {\bibfnamefont {B.}~\bibnamefont {Kersevan}},
  \bibinfo {author} {\bibfnamefont {T.}~\bibnamefont {Pierzchala}}, \bibinfo
  {author} {\bibfnamefont {E.}~\bibnamefont {Richter-W\c{a}s}}, \bibinfo
  {author} {\bibfnamefont {Z.}~\bibnamefont {W\c{a}s}},  \emph {et~al.},\
  }\href {\doibase 10.1016/j.cpc.2005.12.018} {\bibfield  {journal} {\bibinfo
  {journal} {Comput. Phys. Commun.}\ }\textbf {\bibinfo {volume} {174}},\
  \bibinfo {pages} {818} (\bibinfo {year} {2006})},\ \Eprint
  {http://arxiv.org/abs/hep-ph/0312240} {arXiv:hep-ph/0312240} \BibitemShut
  {NoStop}%
\bibitem [{\citenamefont {Beringer}\ \emph {et~al.}(2012)\citenamefont
  {Beringer} \emph {et~al.}}]{Beringer:1900zz}%
  \BibitemOpen
  \bibfield  {author} {\bibinfo {author} {\bibfnamefont {J.}~\bibnamefont
  {Beringer}} \emph {et~al.} (\bibinfo {collaboration} {Particle Data Group}),\
  }\href {\doibase 10.1103/PhysRevD.86.010001} {\bibfield  {journal} {\bibinfo
  {journal} {Phys. Rev. D}\ }\textbf {\bibinfo {volume} {86}},\ \bibinfo
  {pages} {010001} (\bibinfo {year} {2012})},\ \bibinfo {note} {and 2013
  partial update for the 2014 edition.}\BibitemShut {Stop}%
\bibitem [{\citenamefont {Brun}\ \emph {et~al.}()\citenamefont {Brun},
  \citenamefont {Bruyant}, \citenamefont {Maire}, \citenamefont {McPherson},\
  and\ \citenamefont {Zanarini}}]{Brun:1987ma}%
  \BibitemOpen
  \bibfield  {author} {\bibinfo {author} {\bibfnamefont {R.}~\bibnamefont
  {Brun}}, \bibinfo {author} {\bibfnamefont {F.}~\bibnamefont {Bruyant}},
  \bibinfo {author} {\bibfnamefont {M.}~\bibnamefont {Maire}}, \bibinfo
  {author} {\bibfnamefont {A.}~\bibnamefont {McPherson}}, \ and\ \bibinfo
  {author} {\bibfnamefont {P.}~\bibnamefont {Zanarini}},\ }\href@noop {} {\
  }\bibinfo {note} {{\tt GEANT} 3.21, CERN Report DD/EE/84-1
  (1987)}\BibitemShut {NoStop}%
\bibitem [{\citenamefont {Sj{\"{o}}strand}(1994)}]{Sjostrand:1993yb}%
  \BibitemOpen
  \bibfield  {author} {\bibinfo {author} {\bibfnamefont {T.}~\bibnamefont
  {Sj{\"{o}}strand}},\ }\href {\doibase 10.1016/0010-4655(94)90132-5}
  {\bibfield  {journal} {\bibinfo  {journal} {Comput. Phys. Commun.}\ }\textbf
  {\bibinfo {volume} {82}},\ \bibinfo {pages} {74} (\bibinfo {year}
  {1994})}\BibitemShut {NoStop}%
\bibitem [{\citenamefont {Lange}(2001)}]{Lange:2001uf}%
  \BibitemOpen
  \bibfield  {author} {\bibinfo {author} {\bibfnamefont {D.}~\bibnamefont
  {Lange}},\ }\href {\doibase 10.1016/S0168-9002(01)00089-4} {\bibfield
  {journal} {\bibinfo  {journal} {Nucl. Instrum. and. Meth. A}\ }\textbf
  {\bibinfo {volume} {462}},\ \bibinfo {pages} {152} (\bibinfo {year}
  {2001})}\BibitemShut {NoStop}%
\bibitem [{\citenamefont {Berends}\ \emph {et~al.}(1986)\citenamefont
  {Berends}, \citenamefont {Daverveldt},\ and\ \citenamefont
  {Kleiss}}]{Berends:1986ig}%
  \BibitemOpen
  \bibfield  {author} {\bibinfo {author} {\bibfnamefont {F.~A.}\ \bibnamefont
  {Berends}}, \bibinfo {author} {\bibfnamefont {P.}~\bibnamefont {Daverveldt}},
  \ and\ \bibinfo {author} {\bibfnamefont {R.}~\bibnamefont {Kleiss}},\ }\href
  {\doibase 10.1016/0010-4655(86)90115-3} {\bibfield  {journal} {\bibinfo
  {journal} {Comput. Phys. Commun.}\ }\textbf {\bibinfo {volume} {40}},\
  \bibinfo {pages} {285} (\bibinfo {year} {1986})}\BibitemShut {NoStop}%
\bibitem [{\citenamefont {Hanagaki}\ \emph {et~al.}(2002)\citenamefont
  {Hanagaki}, \citenamefont {Kakuno}, \citenamefont {Ikeda}, \citenamefont
  {Iijima},\ and\ \citenamefont {Tsukamoto}}]{Hanagaki:2001fz}%
  \BibitemOpen
  \bibfield  {author} {\bibinfo {author} {\bibfnamefont {K.}~\bibnamefont
  {Hanagaki}}, \bibinfo {author} {\bibfnamefont {H.}~\bibnamefont {Kakuno}},
  \bibinfo {author} {\bibfnamefont {H.}~\bibnamefont {Ikeda}}, \bibinfo
  {author} {\bibfnamefont {T.}~\bibnamefont {Iijima}}, \ and\ \bibinfo {author}
  {\bibfnamefont {T.}~\bibnamefont {Tsukamoto}},\ }\href {\doibase
  10.1016/S0168-9002(01)02113-1} {\bibfield  {journal} {\bibinfo  {journal}
  {Nucl. Instrum. and Meth. A}\ }\textbf {\bibinfo {volume} {485}},\ \bibinfo
  {pages} {490} (\bibinfo {year} {2002})},\ \Eprint
  {http://arxiv.org/abs/hep-ex/0108044} {arXiv:hep-ex/0108044} \BibitemShut
  {NoStop}%
\bibitem [{\citenamefont {Abashian}\ \emph
  {et~al.}(2002{\natexlab{b}})\citenamefont {Abashian}, \citenamefont {Abe},
  \citenamefont {Abe}, \citenamefont {Behera}, \citenamefont {Handa} \emph
  {et~al.}}]{Abashian:2002bd}%
  \BibitemOpen
  \bibfield  {author} {\bibinfo {author} {\bibfnamefont {A.}~\bibnamefont
  {Abashian}}, \bibinfo {author} {\bibfnamefont {K.}~\bibnamefont {Abe}},
  \bibinfo {author} {\bibfnamefont {K.}~\bibnamefont {Abe}}, \bibinfo {author}
  {\bibfnamefont {P.}~\bibnamefont {Behera}}, \bibinfo {author} {\bibfnamefont
  {F.}~\bibnamefont {Handa}},  \emph {et~al.},\ }\href {\doibase
  10.1016/S0168-9002(02)01164-6} {\bibfield  {journal} {\bibinfo  {journal}
  {Nucl. Instrum. and Meth. A}\ }\textbf {\bibinfo {volume} {491}},\ \bibinfo
  {pages} {69} (\bibinfo {year} {2002}{\natexlab{b}})}\BibitemShut {NoStop}%
\bibitem [{\citenamefont {Banerjee}\ \emph {et~al.}(2008)\citenamefont
  {Banerjee}, \citenamefont {Pietrzyk}, \citenamefont {Roney},\ and\
  \citenamefont {W\c{a}s}}]{Banerjee:2007is}%
  \BibitemOpen
  \bibfield  {author} {\bibinfo {author} {\bibfnamefont {S.}~\bibnamefont
  {Banerjee}}, \bibinfo {author} {\bibfnamefont {B.}~\bibnamefont {Pietrzyk}},
  \bibinfo {author} {\bibfnamefont {J.~M.}\ \bibnamefont {Roney}}, \ and\
  \bibinfo {author} {\bibfnamefont {Z.}~\bibnamefont {W\c{a}s}},\ }\href
  {\doibase 10.1103/PhysRevD.77.054012} {\bibfield  {journal} {\bibinfo
  {journal} {Phys. Rev. D}\ }\textbf {\bibinfo {volume} {77}},\ \bibinfo
  {pages} {054012} (\bibinfo {year} {2008})},\ \Eprint
  {http://arxiv.org/abs/0706.3235} {arXiv:0706.3235} \BibitemShut {NoStop}%
\bibitem [{\citenamefont {Lefebvre}\ \emph {et~al.}(2000)\citenamefont
  {Lefebvre}, \citenamefont {Keeler}, \citenamefont {Sobie},\ and\
  \citenamefont {White}}]{Lefebvre:1999yu}%
  \BibitemOpen
  \bibfield  {author} {\bibinfo {author} {\bibfnamefont {M.}~\bibnamefont
  {Lefebvre}}, \bibinfo {author} {\bibfnamefont {R.~K.}\ \bibnamefont
  {Keeler}}, \bibinfo {author} {\bibfnamefont {R.}~\bibnamefont {Sobie}}, \
  and\ \bibinfo {author} {\bibfnamefont {J.}~\bibnamefont {White}},\ }\href
  {\doibase 10.1016/S0168-9002(00)00323-5} {\bibfield  {journal} {\bibinfo
  {journal} {Nucl. Instrum. and Meth. A}\ }\textbf {\bibinfo {volume} {451}},\
  \bibinfo {pages} {520} (\bibinfo {year} {2000})},\ \Eprint
  {http://arxiv.org/abs/hep-ex/9909031} {arXiv:hep-ex/9909031} \BibitemShut
  {NoStop}%
\bibitem [{\citenamefont {Barate}\ \emph {et~al.}(1998)\citenamefont {Barate}
  \emph {et~al.}}]{Barate:1997tt}%
  \BibitemOpen
  \bibfield  {author} {\bibinfo {author} {\bibfnamefont {R.}~\bibnamefont
  {Barate}} \emph {et~al.} (\bibinfo {collaboration} {ALEPH Collaboration}),\
  }\href {\doibase 10.1007/s100520050184} {\bibfield  {journal} {\bibinfo
  {journal} {Eur. Phys. J. C}\ }\textbf {\bibinfo {volume} {4}},\ \bibinfo
  {pages} {29} (\bibinfo {year} {1998})}\BibitemShut {NoStop}%
\bibitem [{\citenamefont {Aubert}\ \emph {et~al.}(2009)\citenamefont {Aubert}
  \emph {et~al.}}]{Aubert:2008an}%
  \BibitemOpen
  \bibfield  {author} {\bibinfo {author} {\bibfnamefont {B.}~\bibnamefont
  {Aubert}} \emph {et~al.} (\bibinfo {collaboration} {BaBar Collaboration}),\
  }\href {\doibase 10.1016/j.nuclphysbps.2009.03.034} {\bibfield  {journal}
  {\bibinfo  {journal} {Nucl. Phys. Proc. Suppl.}\ }\textbf {\bibinfo {volume}
  {189}},\ \bibinfo {pages} {193} (\bibinfo {year} {2009})},\ \Eprint
  {http://arxiv.org/abs/0808.1121} {arXiv:0808.1121} \BibitemShut {NoStop}%
\bibitem [{\citenamefont {Antonelli}\ \emph {et~al.}(2013)\citenamefont
  {Antonelli}, \citenamefont {Cirigliano}, \citenamefont {Lusiani},\ and\
  \citenamefont {Passemar}}]{Antonelli:2013usa}%
  \BibitemOpen
  \bibfield  {author} {\bibinfo {author} {\bibfnamefont {M.}~\bibnamefont
  {Antonelli}}, \bibinfo {author} {\bibfnamefont {V.}~\bibnamefont
  {Cirigliano}}, \bibinfo {author} {\bibfnamefont {A.}~\bibnamefont {Lusiani}},
  \ and\ \bibinfo {author} {\bibfnamefont {E.}~\bibnamefont {Passemar}},\
  }\href@noop {} {\  (\bibinfo {year} {2013})},\ \Eprint
  {http://arxiv.org/abs/1304.8134} {arXiv:1304.8134} \BibitemShut {NoStop}%
\bibitem [{\citenamefont {Sj{\"{o}}strand}\ \emph {et~al.}(2006)\citenamefont
  {Sj{\"{o}}strand}, \citenamefont {Mrenna},\ and\ \citenamefont
  {Skands}}]{Sjostrand:2006za}%
  \BibitemOpen
  \bibfield  {author} {\bibinfo {author} {\bibfnamefont {T.}~\bibnamefont
  {Sj{\"{o}}strand}}, \bibinfo {author} {\bibfnamefont {S.}~\bibnamefont
  {Mrenna}}, \ and\ \bibinfo {author} {\bibfnamefont {P.~Z.}\ \bibnamefont
  {Skands}},\ }\href {\doibase 10.1088/1126-6708/2006/05/026} {\bibfield
  {journal} {\bibinfo  {journal} {JHEP}\ }\textbf {\bibinfo {volume} {0605}},\
  \bibinfo {pages} {026} (\bibinfo {year} {2006})},\ \Eprint
  {http://arxiv.org/abs/hep-ph/0603175} {arXiv:hep-ph/0603175} \BibitemShut
  {NoStop}%
\end{thebibliography}%

\end{document}